\documentclass[twocolumn]{aastex631}
\usepackage{float}
\usepackage{amsmath}

\newcommand{\mgii}{Mg {\sc ii}}
\newcommand{\feii}{Fe {\sc ii}}
\newcommand{\oii}{[O {\sc ii}]}
\newcommand{\oiii}{[O {\sc iii}]}
\newcommand{\hbeta}{H$\beta$}
\newcommand{\feiiems}{Fe {\sc ii}$^{\ast}$}

\newcommand{\MsolarPerYr}{$M_{\odot}\ yr^{-1}$}
\newcommand{\kms}{$\mathrm{km\ s^{-1}}$}

\shorttitle{Spatially Resolved CGM at $z \approx$ 0.8}
\shortauthors{A. Shaban et al.}

\begin{document}

\title{Spatially Resolved Circumgalactic Medium around a Star-Forming Galaxy Driving a Galactic Outflow at $z \approx 0.8$}

\correspondingauthor{Ahmed Shaban, Rongmon Bordoloi}
\email{arshaban@ncsu.edu, rbordol@ncsu.edu}

\author[0000-0002-8858-7875]{Ahmed Shaban}
\affil{Department of Physics, North Carolina State University, Raleigh, NC 27695-8202, USA}

\author[0000-0002-3120-7173]{Rongmon Bordoloi}
\affiliation{Department of Physics, North Carolina State University, Raleigh, NC 27695-8202, USA}

\author[0000-0002-7893-1054]{John M. O'Meara}
\affiliation{W. M. Keck Observatory, 65-1120 Mamalahoa Hwy, Kamuela, HI 96743, USA}

\author[0000-0002-7559-0864]{Keren Sharon}
\affiliation{Department of Astronomy, University of Michigan, 1085 South University Ave., Ann Arbor, MI 48109, USA}

\author[0000-0002-1883-4252]{Nicolas Tejos}
\affiliation{Instituto de F\'isica, Pontificia Universidad Cat\'olica de Valpara\'iso, Casilla, 4059 Valpara\'iso, Chile}

\author[0000-0003-0389-0902]{Sebastian Lopez}
\affiliation{Departamento de Astronom\'ia, Universidad de Chile, Casilla 36-D, Santiago, Chile}

\author[0000-0002-7864-3327]{C\'edric Ledoux}
\affiliation{European Southern Observatory, Alonso de C\'ordova, 3107 Vitacura, Casilla 19001, Santiago, Chile}

\author[0000-0003-0151-0718]{L. Felipe Barrientos}
\affiliation{Instituto de Astrof\'isica, Pontificia Universidad Cat\'olica de Chile, Av. Vicu\~na Mackenna 4860, 7820436, Macul Santiago, Chile}

\author[0000-0002-7627-6551]{Jane R. Rigby}
\affiliation{Astrophysics Science Division, Code 660, NASA Goddard Space Flight Center, 8800 Greenbelt Rd., Greenbelt, MD 20771, USA}

\begin{abstract}

We report the small-scale spatial variation in cool ($T\sim 10^4 K$) {\mgii} absorption detected in the circumgalactic medium (CGM) of a star-forming galaxy at $z\approx 0.8$. The CGM of this galaxy is probed by a spatially extended bright background gravitationally lensed arc at $z = 2.76$. The background arc continuously samples the CGM of the foreground galaxy at a range of impact parameters between 54 and 66 kpc. The {\mgii} absorption strengths vary by more than a factor of 2 within these ranges. A power-law fit to the fractional variation of absorption strengths yields a coherence length of 5.8 kpc within this range of impact parameters. This suggests a high degree of spatial coherence in the CGM of this galaxy. The host galaxy is driving a strong galactic outflow with a mean outflow velocity $\approx$ -179 {\kms} and mass outflow rate  $\dot{M}_{out}\geq 64_{-27}^{+31} $ {\MsolarPerYr} traced by blueshifted {\mgii} and {\feii} absorption lines. The galaxy itself has a spatially extended emission halo with a maximum spatial extent of $\approx$ 33 kpc traced by {\oii}, {\oiii}, and {\hbeta} emission lines. The extended emission halo shows kinematic signatures of corotating halo gas with solar metallicity. Taken together, these observations suggest evidence of a baryon cycle that is recycling the outflowing gas to form the next generation of stars.

\end{abstract}


\section{Introduction} \label{sec:intro}
The circumgalactic medium (CGM) around galaxies has emerged as a reservoir of baryons and metals across cosmic time \citep{Chen2010Outflows, Rudie2012CGMHI, bordoloi2014dependence, Peeples2014MetalsCGM, Werk2014COSHalosCGM, Peroux2020CosmicBaryon, Lehner2022CGM, Bordoloi2024EIGER4CGM}. The CGM plays an important role in regulating the growth in stellar mass and quenching of star-formation in galaxies \citep{tumlinson2017circumgalactic}. The chemical enrichment of the CGM can occur via metal-enriched outflows from galaxies, which also provide energy to heat the CGM, thereby playing a major role in galaxy evolution \citep{somerville2015physical, Vogelsberger2020CosmoSimGalaxy, Faucher-Giguere2023CGM,thompson2024theory}. Without galactic outflows, modern galaxy evolution models fail to reproduce the observed population of galaxies \citep{Oppenheimer2006SimulationsOutflows, Somerville2008Semi-analyticModel, hopkins2014galaxies, Vogelsberger2014Illustris, Schaye2015EAGLEsimulation,  somerville2015physical, Naab2017TheoreticalChallenges,Pillepich2018SimulatingGFormation, Faucher-Giguere2023CGM}. 

Observations have shown that in star-forming galaxies, galactic outflows are ubiquitous across cosmic time \citep{Heckman2000GalacticSuperwinds,veilleux2005galactic,Chen2010Outflows, Rubin2010PersistenceWinds,Rubin2014UpiquitousOutflows, bordoloi2014dependence, heckman2015systematic, chisholm2015scaling, chisholm2016shining, rupke2018review, Fox2019MilkyWayOutflows, Veilleux2020CoolOutflows, Clark2022GasFlowsMW, Xu2022CLASSYIII,  KeerthiVasan2023OutflowsCosmicNoon, Xu2023OutflowsJWST}. These outflows are multiphased, consisting of ionized, neutral, and molecular gas \citep{Strickland2009SupernovaFeedbackM82, Thompson2016AnOriginMultiphase,Fluetsch2021MultiPhaseOutflows, ReichardtChu2022DuvetOutflows}. The outflows play a key role in chemical enrichment of the CGM by transporting enriched gas out of the galaxy \citep{muratov2017metalflowsCGM, Hafen2019CGMoriginsFIRE}, setting the mass-metallicity relation \citep{tremonti2004origin, Ma2016MZrelation} and regulating and quenching star-formation in galaxies \citep{Hopkins2010MaximumSFDensity, hopkins2012stellar, Spilker2019qQuench}.

The CGM gas is typically very diffuse, which makes it very hard to detect it in emission. Therefore, it is more common to study the CGM in absorption in the spectra of bright background sources, like quasi-stellar objects (QSO) or galaxies, with some impact parameter from the host galaxy \citep{rubin2010galaxies, bordoloi2011radial,  Bielby2019QSAGE}. The lack of a large number of bright background sources typically restricts such studies of statistical analysis of the CGM. To constrain the structure and cohesion of the CGM, we need to probe the CGM at multiple sight lines. However, having more than one QSO sight line probing a CGM is rare. 

Recently, the introduction of wide-field integral field unit (IFU) spectrographs in 10 m class telescopes, combined with gravitationally lensed galaxies or quasars, has made it increasingly possible to probe the CGM of a single foreground galaxy along multiple sight lines \citep{bordoloi2016spatially, Bordoloi2022,lopez2018clumpy,rubin2018AndromedaParachute, afruni2023CoherenceLensCGM, shaban2023dissecting, Dutta2024LensedQSO}. 

In addition, these IFUs have allowed us to map the diffuse emission of the CGM gas or the densest part of the outflow gas in the CGM \citep{rupke2019100, burchett2020circumgalactic, zabl2021muse,  Leclercq2022MUSEMgII_IGrM, Shaban202230kpc, Dutta2023EmissionHalos, Vasan2024LensedGalaxy, Kusakabe2024MXDFCGM, Perrota2024OIINebulae}. These pioneering efforts have opened up a new avenue of studying the CGM simultaneously in absorption and emission in a spatially resolved manner.

In this paper, we demonstrate this unique capability and simultaneously map the extended CGM around a $z \approx 0.8$ galaxy both in absorption and emission. We use IFU observations of a spatially extended gravitationally lensed arc at $z \approx 2.8$, which serves as a spatially extended background source; and the cool CGM gas around the $z \approx 0.8$ galaxy is detected in the spectra of the lensed arc. The IFU observations of the galaxy also show spatially extended emission illuminating the dense CGM, allowing us to map the metallicity and kinematics of that gas.

The paper is structured as follows: Section \ref{sec:Obs} describes the IFU and photometric observations of the field of view (FOV). Section \ref{sec:methods} outlines the methods used for spectral extraction and modeling, ray tracing with the lens model, and spectral energy distribution (SED) fitting. The analysis and results are presented as follows: emission line analysis in Section \ref{sec:emission_analysis}, ‘down-the-barrel’ spectroscopy in Section \ref{sec:Down_the_Barrel_outflow}, and CGM measurements in Section \ref{sec:CGM_Absorption_Results}. Finally, we summarize our conclusions in Section \ref{sec:conclusions}. Throughout, we adopt a Lambda cold dark matter $\Lambda$CDM cosmology with $H_0 \mathrm{ = 70\ km\ s^{-1}Mpc^{-1} }$ and $\Omega_m = 0.3$. All reported magnitudes are in the AB system.

\section{Observations} \label{sec:Obs}
\begin{figure*}
  \centering
  \includegraphics[width=0.7\linewidth]{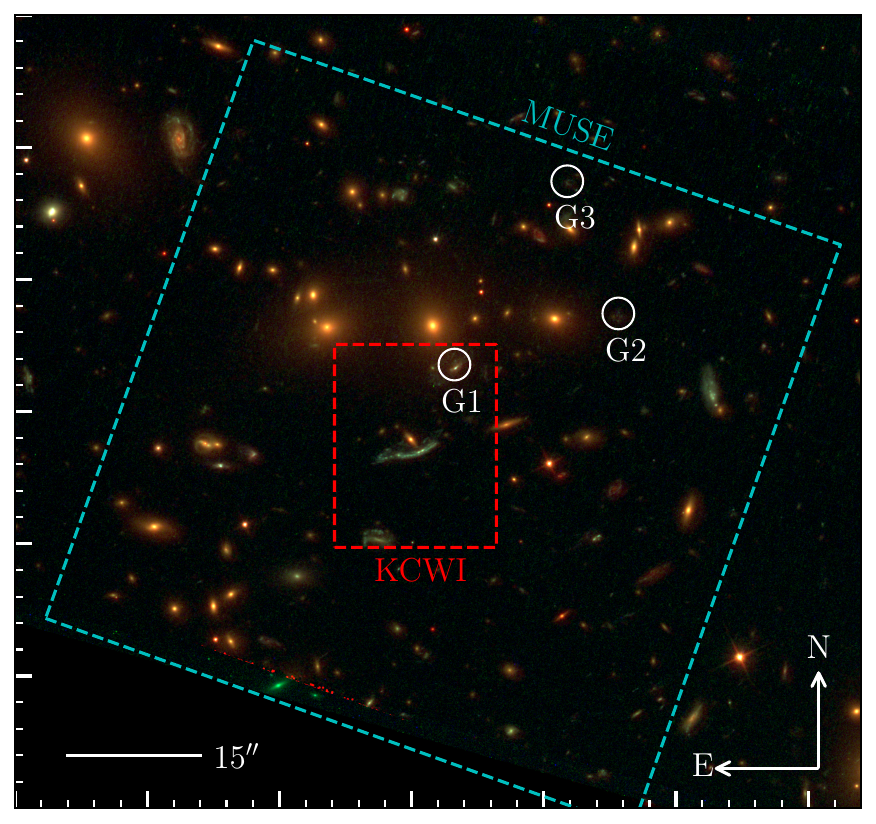}
  \caption{HST red-green-blue (RGB) image of the FOV of the system SGAS J1527. We use WFC3 filters IR-F110W, UVIS-F606W, and UVIS-F475W for the red, green, and blue channels, respectively. The cyan and red dashed polygons mark the boundaries of the Multi Unit Spectroscopic Explorer (MUSE; $60^{\prime\prime}\times 60^{\prime\prime}$) and Keck Cosmic Web Imager (KCWI; $16.5^{\prime\prime}\times 20.4^{\prime\prime}$) IFU spatial coverages, respectively. The three white circles mark three galaxies, which are members of the same group at $z\approx 0.785$. The galaxy G1 is the main target that shows a signature of an outflow. The arc in the middle of the KCWI FOV is of a galaxy at $z\approx 2.76$. }
  \label{fig:HST_MUSE_KCWI_Group}
\end{figure*}

SGAS J152745.1+065219 is a gravitationally lensed galaxy at $z \approx 2.76$ which is distorted to a $6^{\prime\prime}$ extended arc. The strong gravitational lensing is caused by a foreground galaxy cluster at $z \approx 0.392$ \citep{Koester2010TwoLensed}. This arc was discovered as part of the Magellan Evolution of Galaxies Spectroscopic and Ultraviolet Reference Atlas \citep{rigby2018magellanI, rigby2018magellanII}. Figure~\ref{fig:HST_MUSE_KCWI_Group} shows the FOV of this system. In the foreground of this arc, a galaxy is found at $z =0.785273\pm 0.000001$ with an angular separation of $\approx 8^{\prime\prime}$ (Figure~\ref{fig:wl_rgb_kcwi_muse_hst}, cyan square), which is the subject of this study. Throughout this paper, we refer to it as galaxy G1. The background arc provides several intervening lines of sight probing the CGM of the galaxy G1 over 54-66 kpc scales. In the following subsections, we describe the different observations obtained in this field to study the CGM of G1. An interloping foreground galaxy is detected 2$^{\prime\prime}$ east of G1. We extract a spectrum for this galaxy and measure its redshift to be $z_{interloping} = 0.43084\pm 0.00001$. Since it is not associated with G1, we will ignore this galaxy for the remainder of the analysis. In addition, two galaxies G2 and G3 with redshifts $z_{G2} = 0.78494\pm 0.00001$ and $z_{G3} = 0.78542 \pm 0.00001$ to the northwest direction of G1 are detected (Figure~\ref{fig:HST_MUSE_KCWI_Group}). These galaxies are at a separation of 235.7 kpc (G2) and 205.5 kpc (G3) from G1, respectively. G2 and G3 have an average impact parameter from the background arc of 248 kpc and 250 kpc, respectively. All three galaxies G1, G2 and G3 exhibit prominent emission in {\oii}, {\hbeta}, and {\oiii} transitions. Continuum is detected only for G1, where prominent absorption features (e.g., {\mgii} and {\feii} resonant absorption) are detected (see \S~\ref{sec:spectra}). The stellar masses of G2 and G3 are estimated to be $\log (M_\ast / M_{\odot}) $ =   $10.07_{-0.12}^{+0.12}$ and $9.17_{-0.15}^{+0.18}$, respectively, after performing SED fits to the broadband photometry from Hubble Space Telescope (HST) Wide Field Camera 3 (WFC3) observations of the G2 and G3 galaxies (see \S~\ref{sec:properties_photo_sed}). Since G2 and G3 are at very high impact parameters to the background arc relative to G1, we will not consider them for the rest of the analysis. We report their position and properties here for completeness.

\subsection{KCWI and MUSE Integral Field Unit Observations}\label{sec:IFU_Obs}

IFU spectroscopy observations with Keck Cosmic Web Imager \citep[KCWI;][]{morrissey2018keck} were obtained for this field (Program IDs: N080 and K338) in 2018 and 2019, respectively. The total exposure time is 3.56 hr for both observations (N080: 1.56 hr, K338: 2 hr). We refer the reader to \citet{Bordoloi2022} for full details of the observations and data reduction. Here, we give a summary of the observational details. Observations were obtained with the BL grating and the medium IFU slicer, giving a total FOV of $16^{\prime\prime}.5\times 20^{\prime\prime}.4$, and a spectral resolution $R= 1800$. These observations have a wavelength coverage of 350--550 nm. The average seeing during the observing period was $\approx 0^{\prime\prime}.8$. Data were reduced using the KCWI data reduction pipeline \texttt{KCWI\_DRP}\footnote{https://kcwi-drp.readthedocs.io/}\citep{Neill2023KCWIDRPPython} and individual frames were astrometry corrected, trimmed, and co-added after resampling the individual spaxels to a square spaxel grid using the \texttt{kcwitools} package \citep{omeara2022kcwitools}. Frames were resampled using \texttt{Montage}\footnote{http://montage.ipac.caltech.edu/} \citep{Jacob2010Montage, Jacob2010MontageCode}.

Publicly available Very Large Telescope (VLT) MUSE \citep{bacon2010muse} observations are obtained for this field from the ESO archive (Program ID: 0103.A-0485(B), PI: S. Lopez) \citep{Lopez2024CGMabsorptionCIV,Berg2024HIGravitionalArcs}. The MUSE observations were taken in 2019 with a total exposure time of $\approx$ 4 hours. The seeing during observations was in the range of 0.82$^{\prime\prime}$--0.87$^{\prime\prime}$. MUSE has a wavelength coverage of 459.96 -- 935.08 nm (extended range) and a spectral resolution of $R = \frac{\lambda}{\Delta \lambda} \approx 2989$ at the central wavelength for this observation. The mode of observation was Wide Field Mode (WFM) with Adaptive Optics (AO) covering a field of view of $60^{\prime\prime} \times 60^{\prime\prime}$, with a spatial pixel size of $0.2^{\prime\prime}\times 0.2^{\prime\prime}$. The data cube is reduced using the MUSE public data reduction pipeline \citep{Weilbacher2016MUSEPipeline, Weilbacher2020MUSEPipeline}. In the resulting data cube, the observed wavelength values are measured in air. Therefore, we perform air-to-vacuum correction for the wavelength values. Upon initial inspection of the data cube, we notice the presence of some strong skylines in the spectra. To remove these skylines, we use the Zurich Atmospheric Purge (ZAP) algorithm to clean the data cube from their contribution \citep{soto2016zap}. We create a source mask using \texttt{photutils}\footnote{https://photutils.readthedocs.io/en/stable/} \citep{bradley2023photutils}, to identify all bright sources in the field. This mask was fed into ZAP before performing the sky subtraction step. This creates the final sky-subtracted MUSE data cube used in this analysis.

Both the KCWI and MUSE data cubes are astrometry corrected to ensure that all observations have accurate and consistent spatial information. We perform this astrometric correction with a custom script that is described as follows. The ground-based observations are astrometry corrected to match the WCS solution of the HST observations (see \S \ref{sec:photo_obs} for HST observations description). First white light images are created for both KCWI and MUSE data cubes. Then, using \texttt{photutils}, bright sources are detected in both the white light images and the HST image. At least three common bright sources are identified in each image, and the astrometric difference between them with respect to the HST image is calculated. We find an astrometric offset in the KCWI cube coordinates with respect to the space-based HST observations to be $\Delta \alpha = 0.38^{\prime\prime}$ in right ascension ($\alpha$), and $\Delta \delta = 1.34^{\prime\prime}$ in declination ($\delta$), respectively. Similarly, we find an astrometric offset of $\Delta \alpha = 0.82^{\prime\prime}$ and $\Delta \delta = 2.14^{\prime\prime}$ in the MUSE data cube. These offsets are applied as astrometric corrections to each IFU data cube separately.

\begin{figure*}[ht]
  \centering
  \includegraphics[width=\linewidth]{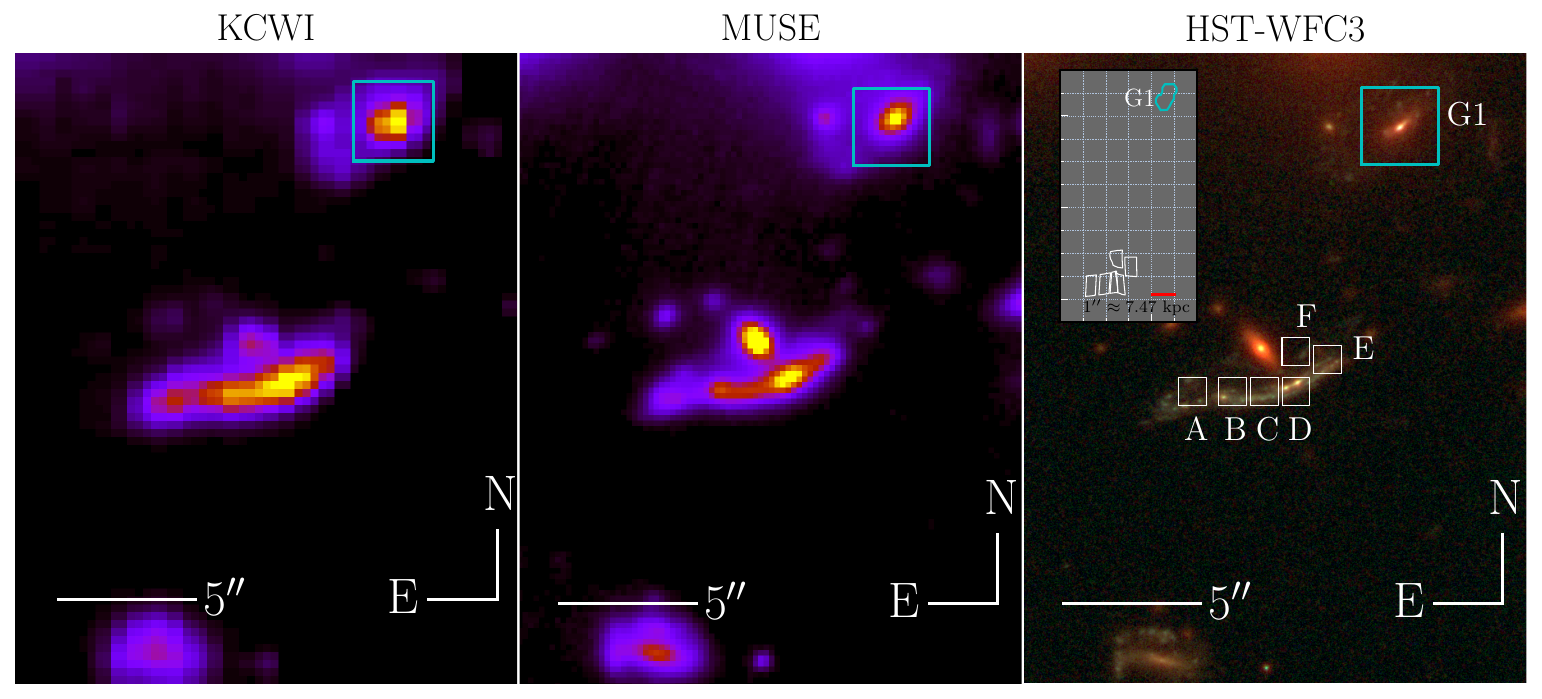}
  \caption{FOV centered on the gravitationally lensed arc SGAS J1527 at $z = 2.76$. The galaxy of interest, G1, at $z = 0.785273$ is marked with a cyan square. White-light images from KCWI observations (left panel) and MUSE observations (middle panel) are shown. The six apertures (each is $1^{\prime\prime} \times 1^{\prime\prime}$) across the arc, where individual 1D spectra are extracted, are marked on the HST/WFC3 image (right panel). The distances from the center of the galaxy to the arc are in the range of 54--66 kpc (see \S \ref{sec:lens_model}). The small light source to the east of G1 is an interloping foreground galaxy at $z \approx 0.431$. The inset in the right panel is the ray-traced locations of G1 and the apertures along the arc in the plane of G1. The grid squares have a size of $1^{\prime\prime}\times 1^{\prime\prime}$, where $1^{\prime\prime} \approx 7.47$ kpc at the redshift of G1. The angular and physical separations between the apertures along the arc and G1 are presented in Table~\ref{tab:apertures_measurements}.}
  \label{fig:wl_rgb_kcwi_muse_hst}
\end{figure*}

Figure~\ref{fig:wl_rgb_kcwi_muse_hst} shows the astrometry-matched observations from KCWI (left panel), MUSE (middle panel) as white-light images, and from HST as an RGB image (right panel). The galaxy of interest in this work (G1) at $z = 0.785273$ (cyan square) and the lensed arc of the background galaxy at $z \approx 2.76$ are marked in these panels of Figure~\ref{fig:wl_rgb_kcwi_muse_hst}. We sum the flux along the wavelength axis for KCWI and MUSE cubes to obtain 2D surface brightness white-light images (see \citealt{Shaban202230kpc} for details).

\begin{deluxetable*}{ccccccc}
\tablecolumns{7}
\tablewidth{0pt}
\tablecaption{Coordinates and properties of the galaxy G1 and apertures across the SGAS J1527 arc.}
\label{tab:apertures_measurements}
\tablehead{
\colhead{Aperture}& 
\colhead{$\alpha$\tablenotemark{a}}& 
\colhead{$\delta$\tablenotemark{a}} & 
\colhead{$\Delta \theta_{G1,i}$  \tablenotemark{b}} & 
\colhead{$D_{G1,i}$  \tablenotemark{c}} &  
\colhead{$\mathrm{W_{r, Mg II 2796}}$ \tablenotemark{d}} &
\colhead{$v_{\mathrm{cen, Mg II2796}}$} \tablenotemark{e}\\
\colhead{ }& 
\colhead{J2000}& 
\colhead{J2000} & 
\colhead{[$^{\prime\prime}$]} & 
\colhead{ [kpc] } &  
\colhead{[\AA]} &
\colhead{[$\mathrm{km\ s^{-1}}$]}
}
\colnumbers
\startdata
G1 & 15:27:44.851 & +06:52:29.076 & -- & -- & $1.97_{-0.06}^{+0.06}$ & -- \\ 
A & 15:27:45.360 & +06:52:19.359 & 8.8 & $65.7_{-1.1}^{+1.0}$ & $0.33_{-0.07}^{+0.08}$ & $25_{-15}^{+15}$ \\ 
B & 15:27:45.262 & +06:52:19.359 & 8.5 & $63.5_{-1.1}^{+1.2}$ & $0.64_{-0.06}^{+0.06}$ & $12_{-8}^{+8}$ \\ 
C & 15:27:45.184 & +06:52:19.359 & 8.3 & $62.1_{-1.1}^{+1.5}$ & $0.48_{-0.05}^{+0.06}$ & $-11_{-9}^{+9}$ \\ 
D & 15:27:45.105 & +06:52:19.359 & 8.3 & $62.0_{-1.3}^{+1.6}$ & $0.36_{-0.04}^{+0.04}$ & $-31_{-8}^{+8}$ \\ 
E & 15:27:45.027 & +06:52:20.524 & 7.4 & $55.0_{-1.1}^{+1.4}$ & $0.29_{-0.06}^{+0.06}$ & $-26_{-15}^{+15}$ \\ 
F & 15:27:45.105 & +06:52:20.816 & 7.2 & $53.9_{-1.0}^{+1.1}$ & $0.30_{-0.07}^{+0.08}$ & $-40_{-26}^{+27}$ \\
\enddata
\tablenotetext{a}{All coordinates are in image plane.}
\tablenotetext{b}{Angular separation from G1 is measured in the plane of G1.}
\tablenotetext{c}{Physical separation from G1 is measured in the plane of G1.}
\tablenotetext{d}{Rest-frame equivalent width of {\mgii} 2796 absorption}
\tablenotetext{e}{Centroid velocity of the Mg II 2796 absorption line with respect to the systemic redshift of G1.}
\vspace{-0.2cm}
\end{deluxetable*}

\subsection{Photometric Observations}\label{sec:photo_obs}
This field has been imaged with the Hubble Space Telescope (HST)- Wide Field Camera 3 (WFC3) (Proposal ID: 13003).These observations are with IR-F160W, IR-F110W, UVIS-F606W and UVIS-F475W, with exposure times 1211s, 1111s, 2400s, and 2400s, respectively. We use the filters F110W, F606W, and F475W to construct an RGB image of the FoV and match the astrometry of the ground-based IFU observations. The HST-WFC3 RGB image is shown in Figure~\ref{fig:HST_MUSE_KCWI_Group} and in the right panel of Figure~\ref{fig:wl_rgb_kcwi_muse_hst}.

The same field has photometry from \textit{Sloan Digital Sky Survey} \citep[SDSS:][]{Fukugita1996SDSSPhoto, Doi2010SDSSPotoResponse} from the filters \textit{u, g, r, i}, and \textit{z}. Each of the observations from these filters has a total exposure of 53.9 seconds. In addition, there are available photometric observations from the \textit{Panoramic Survey Telescope and Rapid Response System} \citep[PanSTARRS:][]{Tonry2012PanSTARRS1Photo, Chambers2016PanSTARRS1Surveys} with filters \textit{g, r, i, z} and \textit{y}. The exposure times of these observations are 853s, 668s, 1530s, 510s, and 780s, respectively. 

We use \texttt{photutils} to estimate the flux in the aperture of G1 in all the filters from HST, SDSS, and PanSTARRS. We also estimate the flux in an empty background region in the field of view without any light sources to calculate the background flux. Then, we subtract this background flux from G1's aperture flux to calculate the AB magnitudes using this background-subtracted flux and the zero point of each filter. We estimate the error on the AB magnitudes of G1 using the standard deviation of the flux in the background aperture times the square root of the area of the G1's aperture. 

We use these photometric AB magnitudes to perform SED fitting to characterize the stellar properties of G1 (see \S \ref{sec:properties_photo_sed}).

\begin{figure*}[ht]
  \centering
  \includegraphics[width=\linewidth]{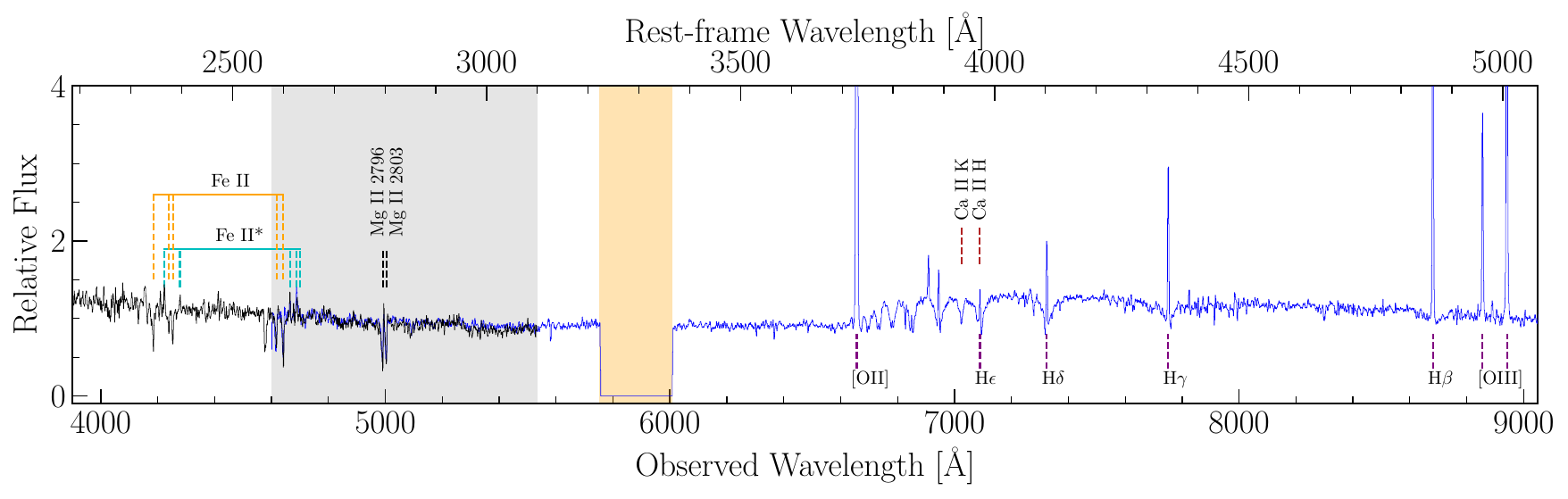}
  \caption{Optimally extracted 1D spectrum of galaxy G1 at $z \approx 0.78$ (cyan square, Figure~\ref{fig:wl_rgb_kcwi_muse_hst}). 1D spectra from KCWI and MUSE data cubes are marked as black and blue solid lines, respectively. The bottom and top x-axes represent the observed and rest-frame wavelengths, respectively. The overlapping wavelength range between the two instruments is marked as the gray vertical band. The wide gap (faint orange band) around $\lambda_{observed} \approx 5800-6000$ {\AA} is the wavelength band of the Sodium D-line laser used in the MUSE Adaptive Optics (AO) mode. The spectra are smoothed with a 3-pixel moving average kernel for presentation. Prominent emission/absorption lines are marked with vertical dashed lines (see Table~\ref{tab:lines_transitions}). }
  \label{fig:full_spectrum_muse_kcwi}
\end{figure*}

\section{Methods} \label{sec:methods}
In this section, we describe the methods used to extract 1D spectra from the IFU data cubes, estimate galaxy properties, model the SED, and ray-trace measurements from the image plane to the source plane of G1.

\begin{deluxetable*}{cccccc}
\tablecolumns{6}
\tablewidth{0pt}
\tablecaption{Prominent absorption and emission lines in the spectrum of the galaxy G1.}
\tablehead{
\colhead{Transition\tablenotemark{a}}&
\colhead{$\lambda$\tablenotemark{b}}&
\colhead{Feature} &
\colhead{$f_0$\tablenotemark{c}} & 
\colhead{KCWI} & 
\colhead{MUSE}
}
\startdata
{\feii} & 2344.212 & Resonant Absorption & 0.1142 & Yes & No \\
 & 2374.460 & Resonant Absorption & 0.0313 & Yes & No \\
 & 2382.764 & Resonant Absorption & 0.320 & Yes & No \\
 & 2586.650 & Resonant Absorption & 0.069125 & Yes & Yes \\
 & 2600.173 & Resonant Absorption & 0.2394 & Yes & Yes \\
{\feiiems}  & 2365.552 & Fine-structure Emission & -- & Yes & No \\
 & 2396.355 & Fine-structure Emission & -- & Yes & No \\
 & 2612.654 & Fine-structure Emission & -- & Yes & Yes \\
 & 2626.451 & Fine-structure Emission & -- & Yes & Yes \\
 & 2632.108 & Fine-structure Emission & -- & Yes & Yes \\
{\mgii} & 2796.351 & Resonant Absorption/Emission & 0.6155 & Yes & Yes \\ 
 & 2803.528 & Resonant Absorption/Emission & 0.3058 & Yes & Yes\\
{\oii}  & 3727.092 & Nebular Emission & -- & No & Yes \\
 & 3729.874 & Nebular Emission & -- & No & Yes \\
H$\epsilon$ & 3971.198 & Emission & -- & No & Yes \\
H$\delta$ & 4102.892 & Emission & -- & No & Yes \\
H$\gamma$ & 4341.692 & Emission & -- & No & Yes \\
H$\beta$ & 4862.708 & Emission & -- & No & Yes \\
{\oiii}  & 4960.295 & Emission & -- & No & Yes \\
{\oiii}  & 5008.239 & Emission & -- & No & Yes \\
\enddata
\vspace{-0.2cm}
\label{tab:lines_transitions}
\tablenotetext{a}{Atomic data from \cite{morton2003atomic} and \cite{leitherer2011ultraviolet}}
\tablenotetext{b}{Vacuum wavelength in angstroms \AA.}
\tablenotetext{c}{Oscillator Strengths}
\end{deluxetable*}

\subsection{Extracting the Spectra}\label{sec:spectra}
We use \texttt{kcwitools}\footnote{https://github.com/pypeit/kcwitools}\citep{omeara2022kcwitools} and \texttt{musetools}\footnote{https://github.com/rongmon/musetools} to read the data cubes and optimally extract 1D spectra from both KCWI and MUSE observations, respectively. Along each aperture, we optimally extract 1D white light-weighted spectrum (see \citealt{Bordoloi2022,shaban2023dissecting} for details). Along the arc, apertures are chosen following \citet{Bordoloi2022} and are $\approx 1^{\prime\prime} \times 1^{\prime\prime}$ in size. The aperture for the galaxy G1 is $\approx 3^{\prime\prime} \times 3^{\prime\prime}$ and shown as a cyan square in Figure~\ref{fig:wl_rgb_kcwi_muse_hst}. This aperture captures the stellar light of G1, as shown in the HST image. The separation distances between G1 and the apertures in the plane of G1 are summarized in Table~\ref{tab:apertures_measurements}.

The 1D extracted spectra for galaxy G1 from KCWI and MUSE are shown in Figure~\ref{fig:full_spectrum_muse_kcwi}. The bluer part of the spectrum extracted from KCWI (black line), shows several resonant absorption features: {\mgii} 2796, 2803 {\AA}, and {\feii} 2344, 2374, 2382, 2586, 2600 {\AA}. Further several fine-structure emission lines: {\feiiems} 2365, 2396, 2612, 2626, and 2632 {\AA} are detected.

In the MUSE spectrum (blue line), a series of prominent emission lines, e.g., {\oii} 3727, 3729 {\AA}, {\oiii} 4960, 5008 {\AA}, and the Hydrogen Balmer Series lines $H\beta$, $H\gamma$, $H\delta$, and $H\epsilon$ are detected. Additionally, resonant absorption lines: {\mgii} 2796, 2803 {\AA}, and {\feii} 2586 {\AA}, 2600 {\AA} lines are detected. Along with these lines, three weak {\feiiems} fine-structure emission lines: {\feiiems} 2612, 2626, and 2632 {\AA} are detected. We summarize the important lines used in this analysis in Table~\ref{tab:lines_transitions}.

The combination of both KCWI and MUSE spectrographs allows simultaneous investigation of the rest frame spectra of galaxy G1 from near-UV to optical ($2000 \leq \lambda/\AA \leq 5200$). The KCWI and MUSE spectrographs overlap in the wavelength range between $\lambda_{rest} =$ 2580 -- 2640 {\AA} ($\lambda_{observed} = $ 4606 -- 5713 {\AA}). In this wavelength range, {\mgii} 2796, 2803 {\AA}, {\feii} 2586, 2600 {\AA}, and {\feiiems} 2612, 2626, 2632 {\AA} are detected in both spectrographs. We leverage this information and simultaneously fit these lines while accounting for different line spread functions (LSF) (see \S \ref{sec:simultaneous_fit}). This allows us to increase the signal-to-noise of the final fits while simultaneously accounting for different resolutions of the two instruments.

In addition to the galaxy, we extract 6 spectra across the lensed arc in Figure~\ref{fig:wl_rgb_kcwi_muse_hst} from both IFU observations. These sightlines probe the CGM of galaxy G1. Each one of these spectra is extracted from a square aperture with size $1^{\prime\prime}\times 1^{\prime\prime}$, shown as white squares in the right panel of Figure~\ref{fig:wl_rgb_kcwi_muse_hst}. We detect {\mgii} 2796, 2803 {\AA} absorption at the redshift of galaxy G1 along these sightlines. The coordinates of G1 and the six apertures across the arc are summarized in Table~\ref{tab:apertures_measurements}. Table~\ref{tab:apertures_measurements} also includes the angular separations between the apertures along the arc and the center of G1, measured in G1's plane.

Before modeling the transitions of interest, each spectrum is continuum-normalized. This is done locally by fitting a 3rd-order polynomial to a 75 {\AA} region around each transition. The flux around this region is divided by the local continuum to obtain the continuum-normalized flux around these transitions. These continuum-normalized spectra are used to perform absorption/emission line strength measurements.

\subsection{Ray-tracing and Lens Model}\label{sec:lens_model}
In order to calculate the distances from G1 to the sightlines along the arc, we need to account for the effect of gravitational lensing due to the foreground galaxy cluster. This galaxy cluster is at redshift $z = 0.392\pm 0.005$ \citep{bayliss2011GeminiLenses}. We follow the ray-tracing procedure described in \cite{Bordoloi2022} and \citet{shaban2023dissecting} to compute the physical separations between the center of G1 and the centers of the six sightlines across the arc in the plane of G1 at $z = 0.785$ using the lens model from \citet{sharon2020strong}. We obtain the physical separations by multiplying the delensed angular separation, from the ray-tracing step, by the angular diameter distance $D_A$ of G1 \citep{hogg1999distance}. The full details of the lens model are described in \citet{Bordoloi2022} and \citet{sharon2020strong}.
\\

\subsection{Spectral Energy Distribution (SED) Fitting}\label{sec:properties_photo_sed}
In order to obtain the stellar properties of the galaxy, we perform SED fitting using the photometric measurements from HST, PanSTARRS, and SDSS. We use \texttt{photutils} to perform aperture photometry to obtain the magnitudes of G1 and the corresponding photometric uncertainties across the different photometric bands (see \S \ref{sec:photo_obs}).

We perform the SED fitting using \texttt{prospector}\footnote{https://prospect.readthedocs.io/en/latest/}\citep{johnson2021propspector}. \texttt{prospector} uses the flexible stellar population synthesis model from \citet{Conroy2009FSPS, Conroy2010FSPS} and its python implementation \texttt{python-FSPS} \citep{Johnson2023PyFSPS}. We assume a simple parametric decaying star-formation history (SFH): $SFH \propto e^{-t_{l}/\tau}$, where $t_{l}$ is the look-back time, and $\tau$ is the characteristic timescale for SFH, which is one of the free parameters of our SED model. The other free parameters for this model are the stellar mass $M_{\ast}$, the dust extinction $A_V$, the age $t_{age}$ of the galaxy at the measured redshift, and the stellar metallicity $Z_{\ast}$. In addition, we use the nebular emission template that has the gas metallicity $Z_{gas}$, and gas ionization parameter $U$ as free parameters to reproduce the emission lines of G1. To sample the posterior distribution of the model parameters, we employ the nested sampling algorithm implemented in \texttt{dynesty}\footnote{https://github.com/joshspeagle/dynesty}, as provided by the \texttt{prospector} framework \citep{Speagle2020Dynesty, Koposov2023dyensty}. We estimate the stellar mass of G1 to be $\mathrm{log(M_{\ast}/M_{\odot}) = 10.84_{-0.10}^{+0.13}}$. From the stellar mass, we estimate the dark matter halo mass of G1 to be $\mathrm{log(M_{halo}/M_{\odot})} = 12.90_{-0.34}^{+0.64}$ \citep{Behroozi2019UniverseMachine}.

\begin{figure*}
  \centering
  \includegraphics[width=\linewidth]{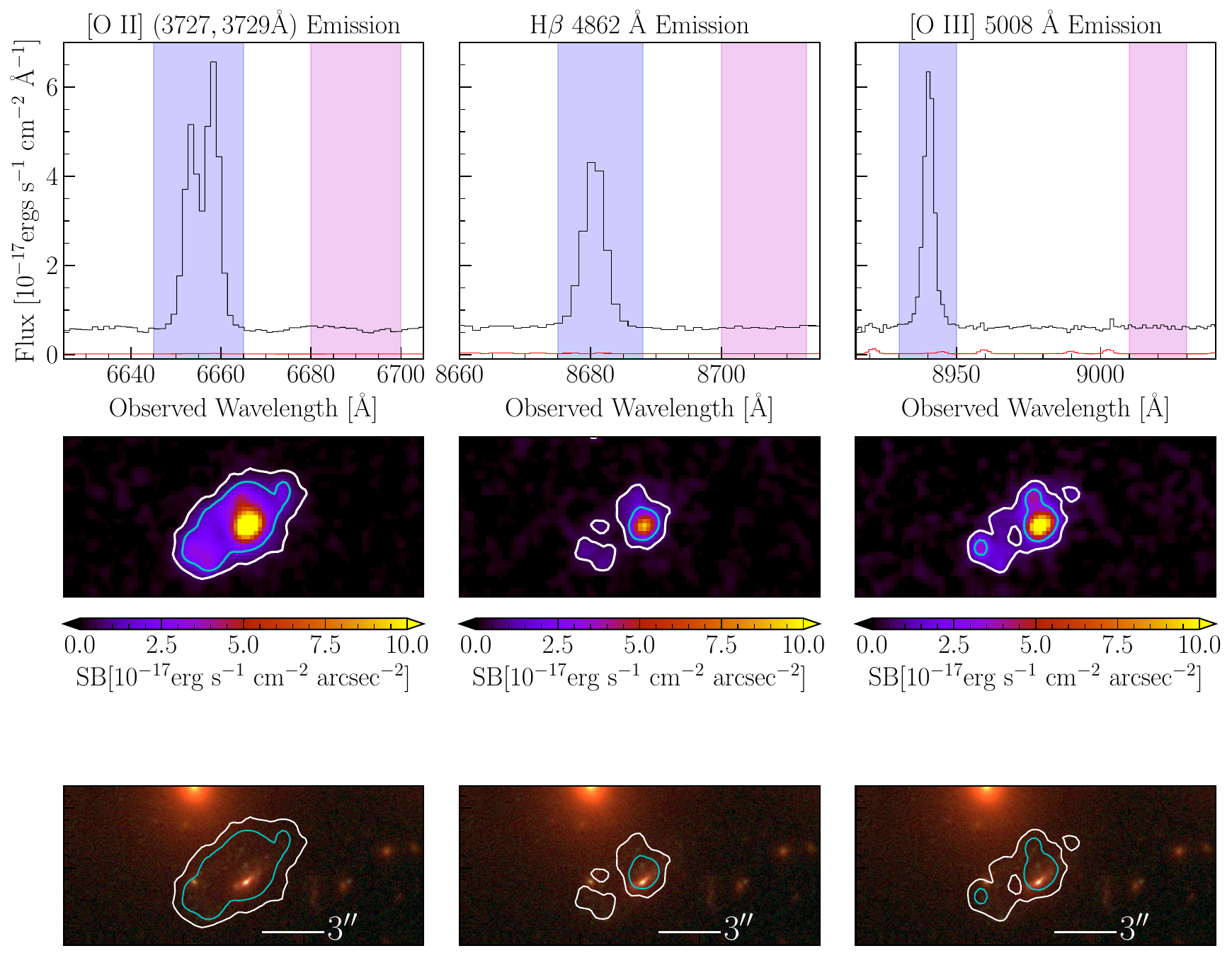}
  \caption{\textit{Top row:} Prominent emission lines in the galaxy: {\oii} 3727, 3729 {\AA} (\textit{Left column}), $H\beta$ (\textit{Middle column}), and {\oiii} 5008 {\AA}(\textit{Right column}) from MUSE, respectively. The solid black and red lines are the flux and the associated uncertainty, respectively. The faint blue and faint magenta windows represent the wavelength windows used to compute the surface brightness maps of the emission+stellar continuum and stellar continuum only, respectively. \textit{Middle row:} The surface brightness maps of the continuum-subtracted emission (blue - magenta from top row) around galaxy G1. The white and cyan contours are the 3 $\sigma$ and $10\sigma$ background surface brightness levels. \textit{Bottom row:} HST/RGB image of the galaxy, with surface brightness contours from the middle row overlaid. The white horizontal $3^{\prime\prime}$ bars in bottom panels corresponds to $\approx$ 10 kpc in the plane of G1.} 
  
  \label{fig:ems_maps_muse}
\end{figure*}

\section{Emission Lines Analysis}\label{sec:emission_analysis}
In this section, we analyze the emission lines detected in the MUSE and KCWI Spectra of the galaxy G1 to estimate the redshift and star formation rate, construct two-dimensional maps of emission line surface brightness, and investivgate the spatial extent and morphology of ionized gas. 

\subsection{Redshift and Star Formation Rate (SFR)}\label{sec:redshift_sfr}
To estimate the redshift ($z$) of galaxy G1, we fit the {\oii} $\lambda\lambda$3727, 3729 doublet using a double Gaussian model and measure the systemic redshift to be $z = 0.785273\pm 0.000001$. The double Gaussian model allows redshift to vary while keeping line ratios between the {\oii} doublet constant. The model was fit using the affine-invariant ensemble sampler \texttt{emcee} \citep{foreman2013emcee}. Similarly, we measure the redshifts of the galaxies G2 and G3 to be $0.78494 \pm 0.00001$ and $0.78542 \pm 0.00001$ using their MUSE spectra, respectively.

Emission lines like {\oii} and H$\beta$ are excellent tracers for measuring star-formation rates (SFR) in galaxies \citep{kennicutt1998star}. Before we measure the fluxes of these lines to calculate the SFR for G1, we need to account for the interstellar dust extinction on G1's spectrum by de-reddening the spectrum. This requires calculating the color excess $E(B-V)$. We calculate $E(B-V)$ using the observed fluxes ratio of the $H\gamma$ and $H\beta$ emission lines and compare it to the intrinsic lines ratio $(H\gamma/ H\beta)_{int} = 0.469$ in case B recombination, which is the standard case in HII regions \citep{osterbrock2006astrophysics}. By measuring the fluxes of the $H\gamma$ and $H\beta$ using a model consisting of a Gaussian and 1D polynomial, we estimate the observed ratio to be $(H\gamma/H\beta)_{obs} = 0.346$. By using attenuation law from \citet{Calzetti2000AttenuationLaw}, which is suitable to describe the effects of dust attenuation in star-forming galaxies, particularly at ultraviolet (UV) and optical wavelengths, we obtain a color excess $E(B-V) = 0.628$ for G1. For de-reddening the spectrum, we use the \texttt{dust\_extinction} module from \texttt{astropy} \citep{astropy2013astropy, astropy2018astropy, astropy2022III} with a dust extinction law from \citet{Fitzpatrick1999Extinction} with total-to-selective extinction ratio of $R_V = 3.1$ for the Milky Way \citep{Cardelli1989Extinction}. From the de-reddened spectrum of G1, we measure the total flux of the H$\beta$ line by fitting a model consisting of a 1D polynomial for the continuum and a Gaussian for the emission line. From the total flux and the angular diameter distance to the galaxy, we obtain the $H\beta$ luminosity $L_{H\beta}$ that is related to $H\alpha$ luminosity $L_{H\alpha} = 2.86 L_{H\beta}$ \citep{osterbrock2006astrophysics}. We use the scaling relation between SFR and $L_{H\alpha}$ from \citet{kennicutt1998star} to obtain an SFR = $30.8\pm 0.5$ {\MsolarPerYr} for G1.

\subsection{Tracing Emission Maps}\label{sec:emission_maps_methods}
To obtain 2D surface brightness images of the emission lines in the G1 spectrum, we follow the procedure described in \cite{Shaban202230kpc}. We select two wavelength windows of equal size: the first window to incorporate the emission line of interest along with the underlying stellar continuum, and the second window to include only the stellar continuum of the galaxy (Figure~\ref{fig:ems_maps_muse}, top panels). Flux from each spatial pixel (spaxel) ($i,j$) in these wavelength windows are summed to obtain the surface brightness $SB_{(i,j)}$ images as follows:
\begin{equation}\label{eq:surface_brightness}
  SB_{(i,j)} =
   \frac{\delta \lambda }{\Delta xy } \sum_{l=\lambda_{min}}^{l=\lambda_{max}} f_{(l, i, j)} ,
\end{equation}
where $f_{(l, i, j)}$ is the flux density at the voxel $(l,i,j)$ in units of $\mathrm{erg\ s^{-1}\ cm^{-2} \AA}$, $\delta\lambda$ is the size of the pixel along the wavelength axis ($\delta \lambda_{MUSE} = 1.25 \AA$, $\delta \lambda_{KCWI} = 1 \AA$), and $\Delta xy$ is the angular area of the spaxel ($\Delta xy_{\ MUSE} = (0.2^{\prime\prime})^2$, $\Delta xy_{\ KCWI} = (0.29^{\prime\prime})^2$).
Finally, the image containing the stellar continuum is subtracted from the emission+continuum image to obtain the continuum-subtracted emission maps. 

\subsection{Spatially Extended Emission} \label{sec:ems_map_results}
Several strong emission lines are detected in the spectrum of galaxy G1 (Figure~\ref{fig:full_spectrum_muse_kcwi}). In this section, we quantify the spatial extent of these strong emission features. Continuum-subtracted surface brightness maps for {\oii}$\lambda\lambda$ 3727, 3729, $H\beta$, and {\oiii}$\lambda$5008 lines are produced as described in \S \ref{sec:emission_maps_methods}. The wavelength windows used to select the emission+stellar continuum regions and the continuum-only regions are shown in the top row of Figure~\ref{fig:ems_maps_muse}. The rest-frame sizes of these wavelength windows are 11.2 {\AA}, 7.3 {\AA}, and 11.2 {\AA} (20{\AA}, 13 {\AA}, and 20 {\AA} in the observed frame) for the {\oii}, {\hbeta}, and {\oiii} emission lines, respectively.

Figure~\ref{fig:ems_maps_muse}, middle row, shows the continuum-subtracted emission maps for these lines. The white and cyan contours represent the $3\sigma$ and $10\sigma$ surface brightness noise levels associated with the maps. The 1$\sigma$ values are computed by selecting an empty region in this FoV with an angular size of $4^{\prime\prime}\times 4^{\prime\prime}$. The $1\sigma$ surface brightness noise level is $1.39 \times 10^{-18}$, $1.83 \times 10^{-18}$, and $2.59 \times 10^{-18}$ $\mathrm{erg\ s^{-1}\ cm^{-2}\ arcsec^{-2}}$ for {\oii}, $H\beta$, and {\oiii} emission maps, respectively. 

Figure~\ref{fig:ems_maps_muse}, bottom row, shows the HST/RGB images of galaxy G1 with the contours from the middle panel overlaid. The {\oii} emission halo at $3\sigma$ level has the maximum physical spatial extent of $\approx$ 33.4 kpc in the plane of G1 at $z \approx 0.8$. The $H\beta$ and {\oiii} emission maps are less spatially extended and show two distinct emission components. This might be because the {\oii} emission lines are the strongest lines in this spectrum. 

The emission maps can be used to measure the mean flux weighted velocity $\langle v \rangle$ (first moment) and the velocity dispersion $\sigma_v$ (second moment) of each spaxel. At each spaxel, a local continuum (first order 1D polynomial) and a Gaussian are fitted around each line of interest. The fitted Gaussian profile is used to compute the $\langle v \rangle$ and $\sigma_v$ for each emission line. Figure~\ref{fig:velocity_moments} shows $\langle v \rangle$ for all spaxels that are greater than 3$\sigma$ detection level. For {\oii} doublet, the kinematics of the {\oii} 3727 line is presented. A clear velocity gradient is seen across the major axis of the galaxy for all three emission lines. We note that the gas distribution is asymmetric, with emission detected along the minor axis of the galaxy. Additionally, the velocity profile shows a skewed distribution, which is inconsistent with a pure rotating disk. Evidence of rotation is also detected in the CGM kinematics of {\mgii} absorbers using QSO and background lensed arc absorption line spectroscopy \citep{Ho2017GasAccretion, Martin2019CGMKinematics,Tejos2021TelltaleCGM}. However, in this case, it is possible that we are detecting a combination of co-rotating halo gas and recycled material from outflows, creating the observed complex kinematic profile. The velocity gradient spans a velocity range of 170 \kms. The velocity dispersion ($\sigma_v$) is of the order of the resolution element of MUSE ($\sim 60-100\ \mathrm{km\ s^{-1}}$) and this work cannot disentangle any variation in $\sigma_v$ in this system. This system is an interesting case with a large co-rotating gas-halo of a star-forming galaxy, which is also driving a strong galactic outflow (see \S \ref{sec:outflow_results})\citep[e.g.,][]{Tejos2021TelltaleCGM}. 

\begin{figure}[ht]
  \centering
  \includegraphics[width=\linewidth]{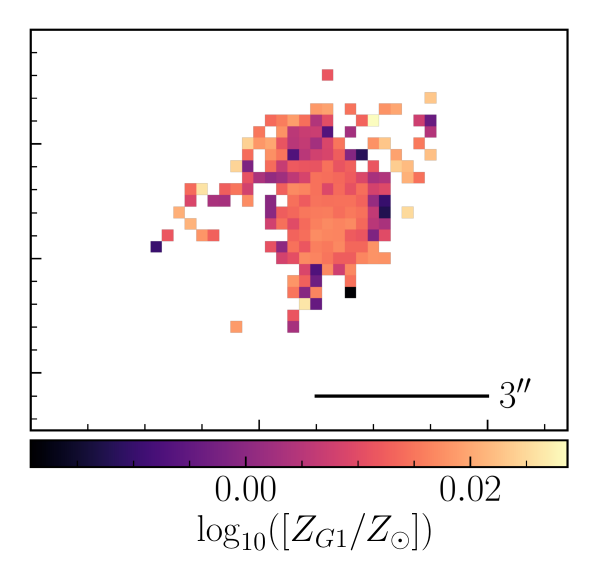}
  \caption{The ratio between the metallicity in each pixel in G1 and the photospheric solar metallicity \citep{Asplund2021SunComposition}. The color bar is log-scale. Most of the pixels are higher than the solar value. The $3^{\prime\prime}$ black line marker corresponds to $\approx$ 10 kpc in the plane of G1.}
  \label{fig:metallicity}
\end{figure}

The spatially extended emission line maps can be used to compute the spatial variation of metallicity across the galaxy. The measured flux of the {\oii} $\lambda$3727, $H\beta$, and {\oiii} $\lambda$4960, 5008 lines are measured using the fitted Gaussians. We use the R23 line ratio: $R23 = ([O II]_{3727} + [O III]_{4960, 5008} ) / H\beta$ \citep{Pagel1979R23Metallcity} and use the calibration between the Oxygen abundance $\mathrm{12 + log(O/H)}$ and $R23$ from \citet{Zaritsky1994metallicity}:

\begin{equation}
\begin{split}
  12 + log\left(\frac{O}{H}\right) & = 9.265 - 0.33\ x \\ 
  & - 0.202\ x^2- 0.207\ x^3 - 0.333\ x^4,
\end{split} 
\end{equation}

\noindent where $x = log(R23)$. As the available observations do not cover other emission lines like Si {\sc ii}, N {\sc ii}, and H$\alpha$, the R23 diagnostic and this calibration represent the current best way to estimate the metallicity from {\oii}, {\oiii}, and $H\beta$ lines \citep{Kewley2002LinesAbundances}. Figure~\ref{fig:metallicity} presents the spatially resolved metallicity map of galaxy G1. The spaxels are color-coded relative to solar metallicity. It is striking that the majority of the spaxels have inferred metallicities higher than solar. We use the light-weighted spectrum of G1 (Figure~\ref{fig:full_spectrum_muse_kcwi}) to measure the average metallicity of galaxy G1 as $\mathrm{12 + log(O/H) = 8.99\pm 0.04}$, which is $1.03\pm 0.01$ times the solar value. The high metallicity in a co-rotating gas-halo and a strong galactic outflow signature in the `down-the-barrel' absorption suggests that this might be a unique case where the `galactic fountain' is seen in action, enriched outflowing gas is falling back in, and co-rotating with the galaxy before losing angular momentum and falling farther into the galaxy \citep{Stewart2013AngularMomentum}. 

\section{Down-the-Barrel Absorption}\label{sec:Down_the_Barrel_outflow}
In this section, we analyze the absorption features of {\mgii} and {\feii} observed in the spectrum of  G1, which exhibit blueshifts relative to the redshift. These down-the-barrel absorption signatures trace outflowing gas along the sight line to G1. We model the absorption spectra to constrain the kinematics, covering fractions, and column densities of the outflowing gas, and derive quantitative estimates of the mass outflow rate.

\subsection{Models of Galactic Outflow}\label{sec:abs_models}
We first describe the models used to quantify the `down-the-barrel' blueshifted absorption in the galaxy G1 seen in {\mgii} and {\feii} absorption lines. We refer the reader to \citet{shaban2023dissecting} for more details of the outflow models. In brief, we model the continuum-normalized flux as:
\begin{equation}\label{eq:absorption_model}
  F(\lambda) = 1 - C_f (\lambda) + C_f(\lambda) e^{-\tau(\lambda)},
\end{equation}
where $C_f(\lambda)$ and $\tau(\lambda)$ are the covering fraction and optical depth as functions of wavelength, respectively \citep{rupke2005outflowsI, sato2009aegis, Rubin2014UpiquitousOutflows}. The optical depth $\tau(\lambda)$ can be expressed as a function of the rest-frame wavelength of each transition ($\lambda_0$) and the optical depth ($\tau_0$) at the center of the absorption line as:
\begin{equation}
  \tau(\lambda) = \tau_0 e^{ (\lambda - \lambda_0)^2 / (\lambda_0 b_D/c)^2 }.
\end{equation}
Here, $b_D$ is the Doppler velocity width, and $c$ is the speed of light in vacuum. Also $\tau_0$ is defined in terms of $\lambda_0$, $b_D$, the column density $N$, and the oscillator strength $f_0$ of the transition as \citep{Draine2011PhysicsISMIGM}:
\begin{equation}
  N[cm^{-2}] = \frac{ \tau_0 b_D [km\ s^{-1}] }{1.497 \times 10^{-15} \lambda_0 [\AA] f_0}.
\end{equation}

We link all the absorption lines of the same ionic transition using the line ratios, which are allowed to vary as free parameters of the model. It is worth noting that the {\mgii} absorption doublet is saturated in G1 spectrum; therefore the total column density of the outflowing gas is strictly a lower limit. For the emission lines, We model each emission line as a Gaussian:
\begin{equation}\label{eq:emission_gaussian}
  F_{ems}(\lambda) = 1 + G(\lambda; A, v_0, b_D),
\end{equation}
where the parameters $A$, $v_0$, and $b_D$ are the amplitude, the central velocity, and the Doppler velocity width of the emission line, respectively. Similar to the absorption lines, we link the amplitudes of the different emission lines of the same species together using the line ratios, which are free parameters of the model \citep{shaban2023dissecting}.

For the absorption and emission lines in galaxy G1, the final model has two components: outflowing ($F_{out}(\lambda)$) and emission ($F_{ems}(\lambda)$, equation \ref{eq:emission_gaussian}). We enforce a \textit{prior} to exclusively consider only the blueshifted velocities as outflowing components (see \citealt{shaban2023dissecting}). The final model is convolved with the line spread function (LSF) of the instrument. Visual inspection of the `down-the-barrel' galaxy G1 spectrum shows that {\mgii} and {\feii} lines have two distinct kinematic components. We account for it by incorporating a second outflowing component to model the high-velocity part of these lines.

\begin{figure*}[ht]
  \centering
  \includegraphics[width=\linewidth]{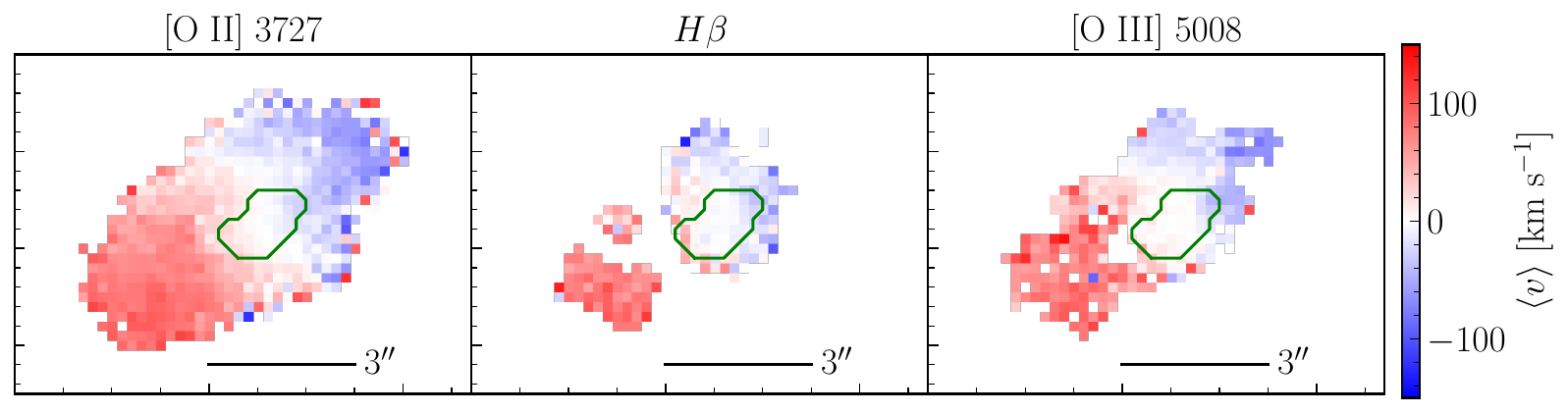}
  \caption{The mean weighted velocity per pixel (first moment) as measured from the emission lines {\oii} 3727 (\textit{left panel}), {$H\beta$} (\textit{middle panel}), and {\oiii} 5008 (\textit{right panel}) for the galaxy G1. The green contour represents the stellar light size of G1 as measured from HST-WFC3 F606W. All the velocity maps show a gradient in velocity from redshifted velocities on the left side to blueshifted velocities on the right side of G1. This is the kinematic signature of a rotating disc. The pixels beyond the green contour are emissions from the gaseous halo around G1. The $3^{\prime\prime}$ black line marker corresponds to $\approx$ 10 kpc in the plane of G1.} 
  \label{fig:velocity_moments}
\end{figure*}

\subsection{Fitting the Models}\label{sec:simultaneous_fit}

The models described above are fitted to the 1D spectra around the transitions of interest using the affine invariant Markov chain Monte Carlo sampler \texttt{emcee} \citep{foreman2013emcee}. This sampler uses Bayesian inference to sample the posterior probability density function (PPDF) of the model parameters. Uniform priors on each model parameter are assumed as described in the previous section. We use a total of 50 walkers, with a total number of 50000 steps. The first 5000 steps are used for the burn-in time. 

Since we have observations from two instruments (MUSE, KCWI), we simultaneously fit the spectra with the same underlying model when both are available. KCWI has a bluer wavelength coverage, therefore it covers {\feii} transitions down to 2344 \AA. MUSE spectra can detect {\feii} transitions down to 2586 \AA. A joint likelihood function is written to account for data from both instruments. In this case, the log-likelihood $log(\mathcal{L})$ can be written as:

\begin{multline}
  log(\mathcal{L(\vec{\theta}})) \propto -\frac{1}{2} \sum_{j=1}^{N_K} \left(\frac{F(\lambda_{K,j}; \vec{\theta}\ )_{K,j} - y_{K,j}}{\sigma_{K,j}} \right)^2  \\ -\frac{1}{2} \sum_{j=N_K+1}^{N_K+N_M} \left( \frac{F(\lambda_{M,j};\vec{\theta}\ )_{M,j} - y_{M,j}}{ \sigma_{M,j} } \right) ^2, 
\end{multline}
where $F(\lambda; \vec{\theta})_K$ and $F(\lambda, \vec{\theta})_M$ are the same physical model for an ionic transition convolved with the LSFs for KCWI and MUSE, respectively. $\vec{\theta}$ are the model parameters, $y_{K,j}$ and $\sigma_{K,j}$ are the $j^{th}$ data points of KCWI flux and flux uncertainty, respectively, $y_{M,j}$ and $\sigma_{M,j}$ are the $j^{th}$ data points of MUSE flux and flux uncertainty, respectively. $N_K$ and $N_M$ represent the total number of data points of the normalized spectrum of the same transition from KCWI and MUSE, respectively.

\begin{figure*}[ht]
  \centering
  \includegraphics[width=\linewidth]{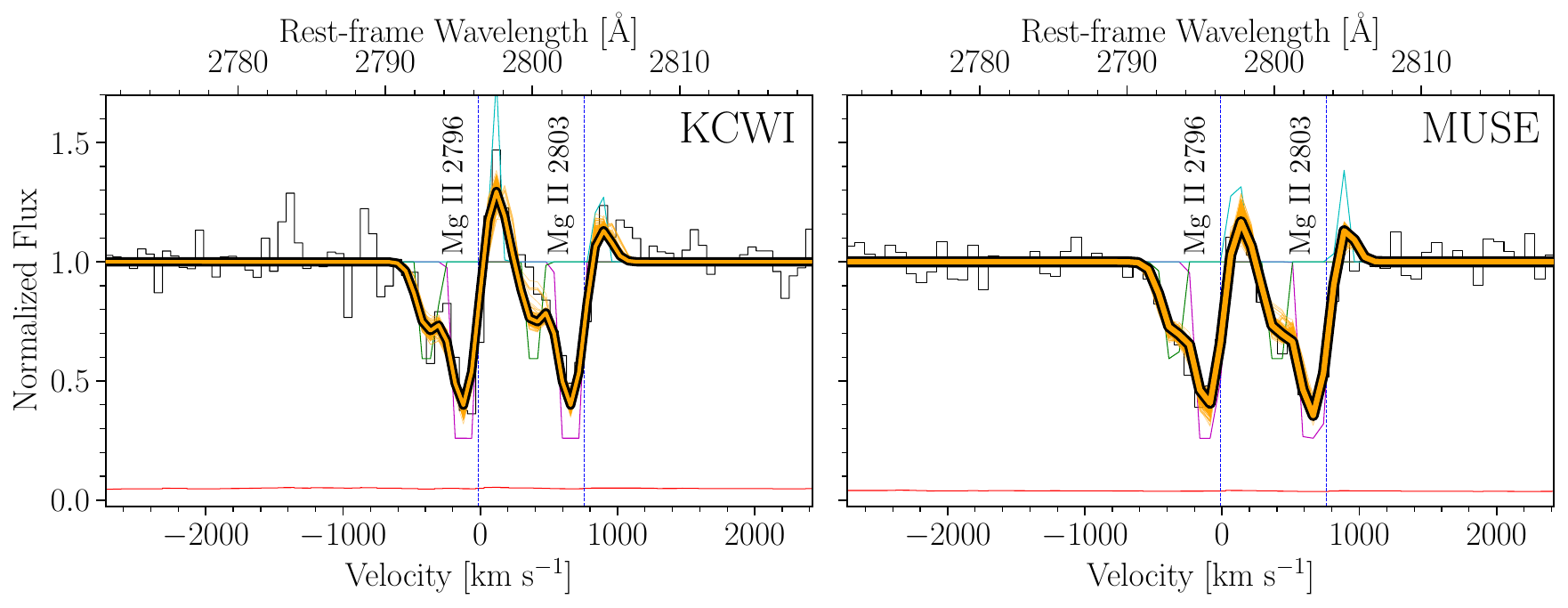}
  \caption{The {\mgii} 2796, 2803 {\AA} doublet in galaxy G1 from KCWI (\textit{Left Panel}) and MUSE (\textit{Right Panel}), respectively. The solid black line and solid red line represent the normalized flux and normalized flux uncertainties. Velocities are centered on the {\mgii} 2796 transition at G1's systemic redshift. The solid dark orange line is the best-fit outflow model convolved with the line spread function (LSF) for each instrument. The faint orange lines are 300 random realizations after sampling the posterior distribution. The purple and green lines for the first and second outflow components of the model. The cyan lines represent the emission component of the model. }
  \label{fig:MgII_KCWI_MUSE_spec_Model}
\end{figure*}

\begin{figure*}[ht]
  \centering
  \includegraphics[width=\linewidth]{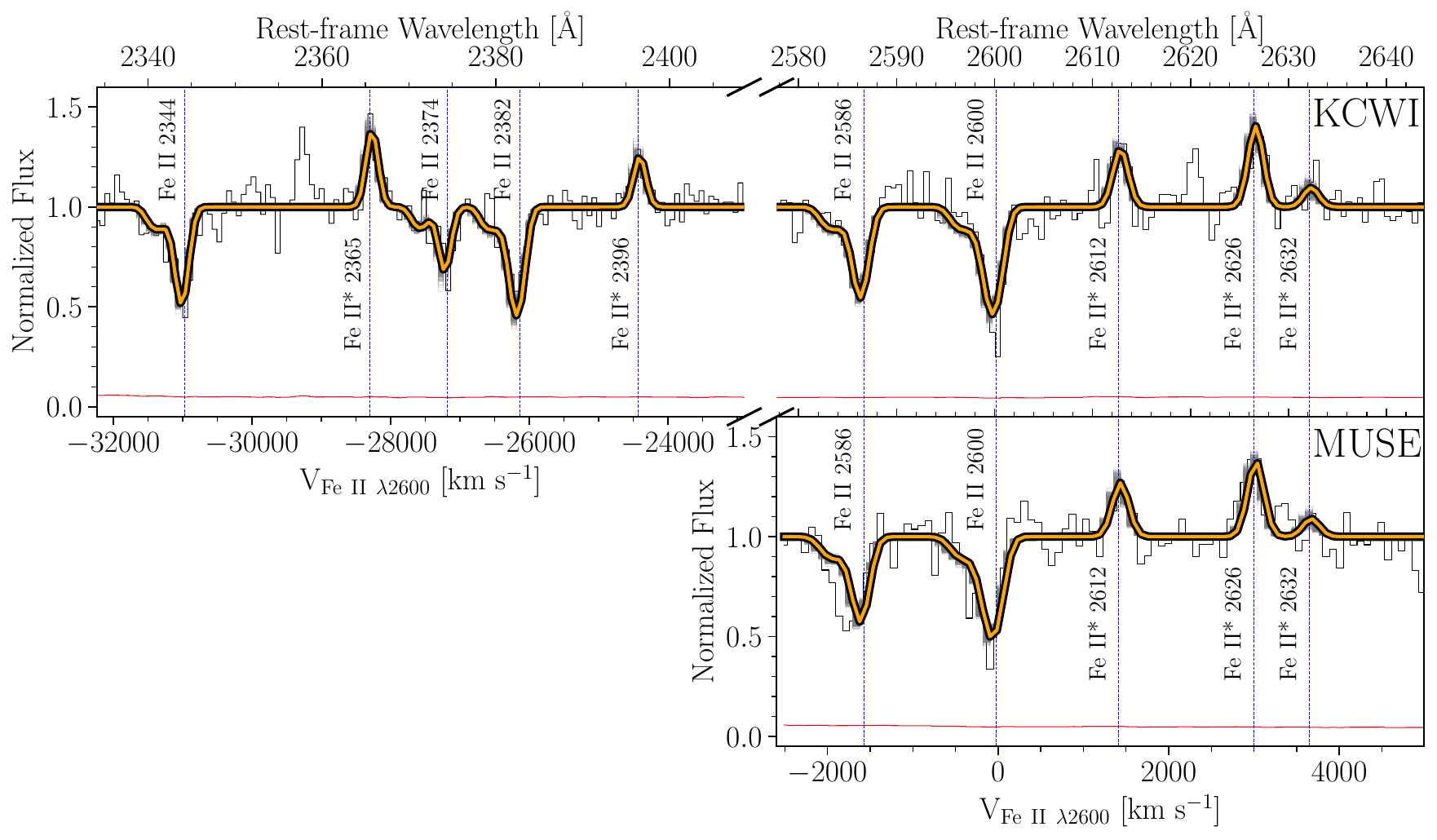}
  \caption{Galactic outflow kinematics traced by {\feii} transitions in galaxy G1 from KCWI (\textit{top row}) and MUSE (\textit{bottom panel}), respectively. All velocities are set to zero with respect to the rest-frame wavelength of {\feii}$\lambda 2600$ at G1's systemic redshift. The solid black and red lines represent the normalized flux and uncertainty. The solid orange line represents the best-fit outflow model. The faint gray lines represent 300 random realizations from the posterior distribution. All the absorption/emission features are marked with dashed vertical blue lines. We observe only five {\feii} transitions lines in the MUSE spectrum because the MUSE wavelength coverage starts at $\lambda_{obs} \approx 4600$ {\AA} ($\lambda_{rest} \approx 2576 $ {\AA}) (see Table~\ref{tab:lines_transitions} for more details). }
  \label{fig:feii_model_spec_kcwi_muse}
\end{figure*}

From the \texttt{emcee} PPDFs realizations of the model parameters of this joint fit, we calculate the rest-frame equivalent width $W_r$, and the weighted outflow velocity $V_{out}$ for each realization as described in \citet{shaban2023dissecting}. The final best-fit results are shown in Figure~\ref{fig:MgII_KCWI_MUSE_spec_Model} and Figure~\ref{fig:feii_model_spec_kcwi_muse}. 

\subsection{Galactic Outflow properties} \label{sec:outflow_results}
The galaxy G1 drives a strong galactic outflow seen as blueshifted {\feii} and {\mgii} absorption lines. We measure these outflow properties and present them in this section.

We use the outflow model highlighted in \S \ref{sec:abs_models} and fit an outflow model with two kinematic components to both the {\mgii} and {\feii} transitions. We first report the outflow kinematics and column densities traced by {\mgii} absorption. Figure~\ref{fig:MgII_KCWI_MUSE_spec_Model} shows the best-fit outflow model (orange line) and 300 realizations of the outflow model (gray lines) created by randomly sampling the posterior distribution are presented. The spectra from KCWI and MUSE are simultaneously fitted, thereby incorporating maximum information into the fitting. 

We use the posterior distribution of the model parameters to measure an absorption weighted mean outflow velocity of $V_{out} = \mathrm{-179_{-5}^{+6}\ km\ s^{-1}}$ from the two components model, where the first component is the lower-velocity component $V_{out,1} = -99_{-8}^ {+11}$\kms, and the second component is the higher-velocity component $V_{out,2} = -367_{-4}^{+5}$\kms. The lower limit on the outflow column density traced by {\mgii} absorption is $\mathrm{log(N_{out}/ cm^{-2})} \geq 15.83_{-0.39}^{+0.23}$, where $N_{out}$ is the sum of the column densities of the two individual outflowing components ($N_{out} = N_{out, 1} + N_{out, 2}$). This corresponds to a rest frame equivalent width of the {\mgii} 2796 line as $W_r = 1.97_{-0.06}^{+0.06}$ {\AA}. The two outflow components have a gas covering fraction of $C_{f,1} = 0.74_{-0.03}^{+0.06}$ for the first component and $C_{f,2} = 0.41_{-0.03}^{+0.03}$ for the second component, respectively. The difference in $C_f$ values highlights that the outflowing gas at the highest velocities is less covered by the light from the background galaxy G1. This suggests that such high-velocity gas is more patchy and is potentially probing gas much farther away than low-velocity outflowing gas \citep[e.g., ][]{chisholm2018feeding}. However, since {\mgii} 2796 line is the strongest absorption line reported in this work (and is not contaminated by emission filling), it is ideal to trace the highest velocity wing of the outflowing gas. We find that galaxy G1 has a maximum outflow velocity $V_{95} = -415_{-7}^{+10}$ \kms, which is the $95^{th}$ percentile velocity from the cumulative distribution function (CDF) of the resulting outflow model \citep{shaban2023dissecting}.

The wavelength ranges of KCWI and MUSE provide mutual coverage of several {\feii} transitions. We simultaneously fit a two kinematic component outflow model to {\feii} 2344, 2374, 2382, 2586, and 2600 absorption lines along with five {\feii} emission lines as described in \S \ref{sec:spectra} and Table~\ref{tab:lines_transitions}. The best-fit outflow model (orange line) and 300 realizations of the outflow model (gray lines) created by randomly sampling the posterior distribution are presented in Figure~\ref{fig:feii_model_spec_kcwi_muse}.

Given the relatively small oscillator strengths ($f_0$) of the {\feii} 2344, 2374, and 2586 absorption transitions in the spectrum of G1, these lines provide more reliable column density estimates compared to the higher-$f_0$ {\feii} transitions or the saturated {\mgii} doublet. However, due to the limited resolution of MUSE and KCWI ($R\sim 2000-3000$), we consider these column densities to be lower limits. From the posterior distribution of the model parameters, we measure the absorption-weighted mean outflow velocity traced by {\feii} absorption as $\mathrm{V_{out} = -122_{-8}^{+8}\ km\ s^{-1} }$. The total {\feii} absorption column density is $\mathrm{log(N_{out}/ cm^{-2}) = 15.82_{-0.23}^{+0.16}}$. This corresponds to a rest frame equivalent width of the {\feii} $\lambda$2382 line of $W_r = 1.58\pm0.05${\AA}. The gas covering fraction of the first kinematic (lower-velocity, $V_{out,1} = -39_{-4}^{+3}$ \kms) component is $C_{f,1} = 0.59_{-0.03}^{+0.03}$ and the second (higher-velocity, $V_{out,2} =-366_{-5}^{+2}$ \kms) component is $C_{f,2} = 0.12 \pm 0.01$, respectively. Similar to the {\mgii} absorption, the gas covering fraction of the high-velocity second component is smaller than the first component. Both the {\mgii} and {\feii} absorptions trace the cool $10^4$K outflowing gas and the highest velocity parts of the outflow are optically thin with a very low covering fraction. The maximum $95^{th}$ percentile outflow velocity traced by the {\feii} 2382 is $V_{95} = -504_{-24}^{+23}$ \kms .
The measured outflow kinematics and column densities of the {\feii} and {\mgii} absorption are summarized in Table~\ref{tab:outflow_measurements}.

\begin{deluxetable*}{ccccccccc}
\tablecolumns{9}
\tablewidth{0pt}
\tablecaption{Measured outflow properties from the {\feii} and {\mgii} transitions of the galaxy G1: (2) is the rest-frame equivalent width, (3) is the mean outflow velocity, (4) is the 95$^{th}$ percentile outflow velocity, (5) and (6) are the column density of the first and second outflow components, (7) and (8) are the covering fractions for the first and second outflow components, and (9) is the measured mass outflow rate. } 
\tablehead{
\colhead{Species}& 
\colhead{$\mathrm{W_{r,out}}$\tablenotemark{a} [\AA]}& \colhead{$V_{out}$\tablenotemark{a} [$\mathrm{km\ s^{-1}}$]} &
\colhead{$V_{95}$\tablenotemark{a} [$\mathrm{km\ s^{-1}}$]} &
\colhead{$\mathrm{log_{10}(N_{1}/cm^{-2})}$ } & 
\colhead{$\mathrm{log_{10}(N_{2}/cm^{-2})}$} &  
\colhead{$C_{f,1}$} &  
\colhead{$C_{f,2}$} & 
\colhead{$\dot{M}_{out}$ [$M_{\odot}\ yr^{-1}$]}
}
\colnumbers
\startdata
Fe II & $1.58_{-0.05}^{+0.05}$ & $-121_{-8}^{+7}$ & $-504_{-24}^{+23}$ & $15.01_{-0.08}^{+0.10}$ & $15.74_{-0.30}^{+0.19}$ & $0.59_{-0.03}^{+0.03}$ & $0.12_{-0.01}^{+0.01}$ & $64_{-27}^{+31}$\\ 
Mg II & $1.97_{-0.06}^{+0.06}$ & $-179_{-4}^{+5}$ & $-415_{-7}^{+10}$ &$15.19_{-0.55}^{+0.57}$ & $15.60_{-0.52}^{+0.29}$ & $0.74_{-0.03}^{+0.06}$ & $0.41_{-0.03}^{+0.03}$ & $248_{-147}^{+170}$\\ 
\enddata
\vspace{-0.2cm}
\label{tab:outflow_measurements}
\tablenotetext{a}{Both quantities are measured for the lines with the highest oscillator strength Fe II $\lambda$2382 and  Mg {\sc II}  $\lambda$2796.}
\end{deluxetable*}

We use the measured outflow gas column density $N$, covering fraction $C_f$, and mean outflow velocity $V_{out}$ to quantify the mass outflow rate ($\dot{M}_{out}$) from galaxy G1. Assuming a thin spherical shell geometry, $\dot{M}_{out}$ can be written as (see \citet{shaban2023dissecting} for derivation): 
\begin{equation}
  \dot{M}_{out} \gtrapprox 22\ \mathrm{M_{\odot}\ yr^{-1}} \frac{C_f}{1.0} \frac{\Omega_w}{4\pi} \frac{V_{out}}{300\ \mathrm{km\ s^{-1}}} \frac{N_{H}}{10^{20} \mathrm{cm^{-2}}} \frac{ R }{5\ \mathrm{kpc}},
\end{equation}
where $N_H$ is the Hydrogen column density, and $R$ is the radius of the outflow. We use the absorption-weighted mean outflow velocity $V_{out}$ and use the mean absorption-weighted gas covering fraction of the two kinematic components as the covering fraction of the outflowing gas $C_f$. We calculate the total gas column density $N_H$ from the total column densities of {\feii} and {\mgii}. To perform this conversion, we assume no-ionization correction, and use the mean metallicity of galaxy G1 as the metallicity of the outflowing gas (\S \ref{sec:ems_map_results}). We also use dust depletion of -1.3 dex and -1.0 dex for {\mgii} and {\feii}, respectively, \citep{jenkins2009unified} and assume solar abundances from \citet{lodders2003solar}.

The radius $R$ of the outflow is one of the least constrained quantities in the calculation of $\dot{M}_{out}$. Since we do not detect any {\mgii} emission above the noise level, we assume a conservative estimate of the mean radius of the outflow to be $R = 5$ kpc \citep[see the appendix of][]{shaban2023dissecting}. At this radius, we get an estimate of the mass outflow rate for G1 at $z = 0.785273$ to be $\dot{M}_{out}(R=5\ \mathrm{kpc}) \gtrapprox 248_{-147}^{+ 170} \ \mathrm{M_{\odot}\ yr^{-1}} $ using {\mgii} measurements. This high uncertainty on the value of $\dot{M}_{out}$ is due to the fact that the {\mgii} lines are saturated. This leads to high uncertainty on column density that gets propagated to the value of $\dot{M}_{out}$.

Given that, we detect optically thin {\feii} lines that better constrain the column density of the outflowing gas. We measure the mass outflow rate of G1 at 5 kpc to be $\dot{M}_{out}(R=5\ \mathrm{kpc}) \gtrapprox 64_{-27}^{+31}\ \mathrm{M_{\odot}\ yr^{-1}} $. Given all these assumptions, our estimates on $\dot{M}_{out}$ are lower limits. In reality, the total mass outflow will be higher than our estimates because we are only accounting for the cool phase $10^4 K$ gas that is traced by {\feii} and {\mgii}. 

By comparing this mass outflow rate from {\feii} to the SFR from {\hbeta} (see \S \ref{sec:redshift_sfr}) of the galaxy G1, we constrain the mass loading factor ($\eta = \dot{M}_{out}/ SFR$) to be $\eta \geq 2.1_{-0.9}^{+1.0}$.

\section{CGM Absorption} \label{sec:CGM_Absorption_Results}

In this section, we report the variation of {\mgii} absorption strengths in the CGM of galaxy G1. The CGM is sampled six times by six lines of sight towards the background arc SGAS1527 between 54 and 66 kpc.
For the six sightlines towards the lensed arc at $z\approx 2.76$, we detect {\mgii} absorption associated with galaxy G1 at $z\approx 0.8$. These are modeled with double Gaussians where the line ratios between the two Gaussian components are between 1-2, and the line separation between them is set at 786 $\mathrm{km\ s^{-1}}$.

\begin{figure*}[ht]
  \centering
  \includegraphics[width=\linewidth]{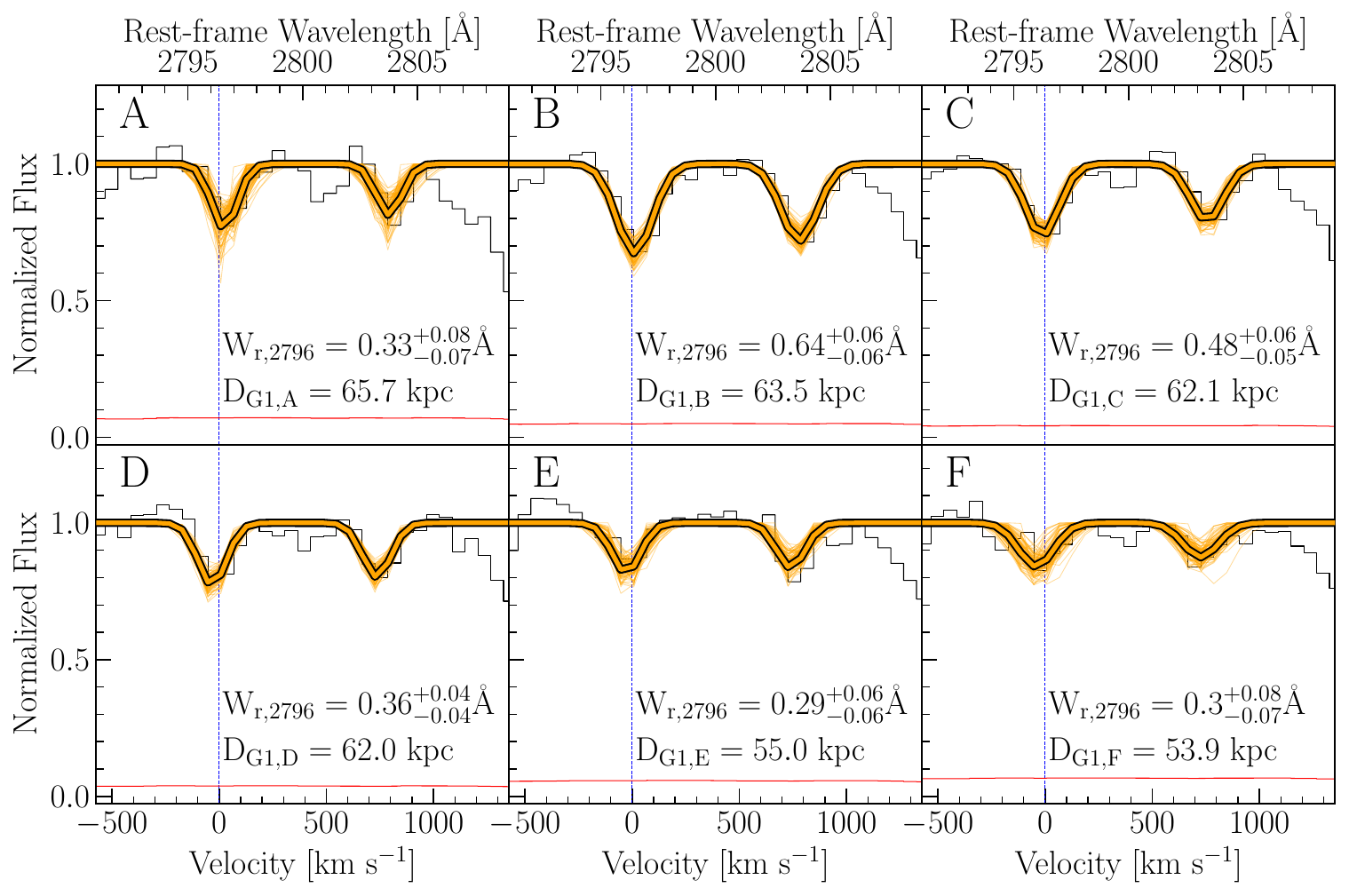}
  \caption{The {\mgii} 2796, 2803 {\AA} absorption in the CGM of the galaxy G1 along 6 sightlines probed by the arc from KCWI as shown in Figure~\ref{fig:wl_rgb_kcwi_muse_hst}.
  Velocities are centered on the {\mgii} 2796 transition at the systemic redshift of galaxy G1, marked as a blue dashed line. The best-fit models for {\mgii} doublet (orange line) and the {\mgii} 2796 rest-frame equivalent width ($W_r$) for each aperture is shown. The faint orange lines are 100 random realizations of the model from the PPDF resulting from \texttt{emcee} runs.}
  \label{fig:cgm_spec_kcwi_muse}
\end{figure*}

Figure~\ref{fig:cgm_spec_kcwi_muse} shows the {\mgii} absorption detected along the six sightlines towards the background arc. The best-fit {\mgii} model (thick orange lines) and the corresponding 300 model realizations (light orange lines) created by randomly sampling the posterior distribution are presented. The {\mgii} 2796 absorption equivalent width ($W_r$) is computed by integrating the best-fit model along each line of sight. Individual {\mgii} absorption strengths are presented in each panel of Figure~\ref{fig:cgm_spec_kcwi_muse}. 

\begin{figure*}
  \centering
   \includegraphics[width=\linewidth]{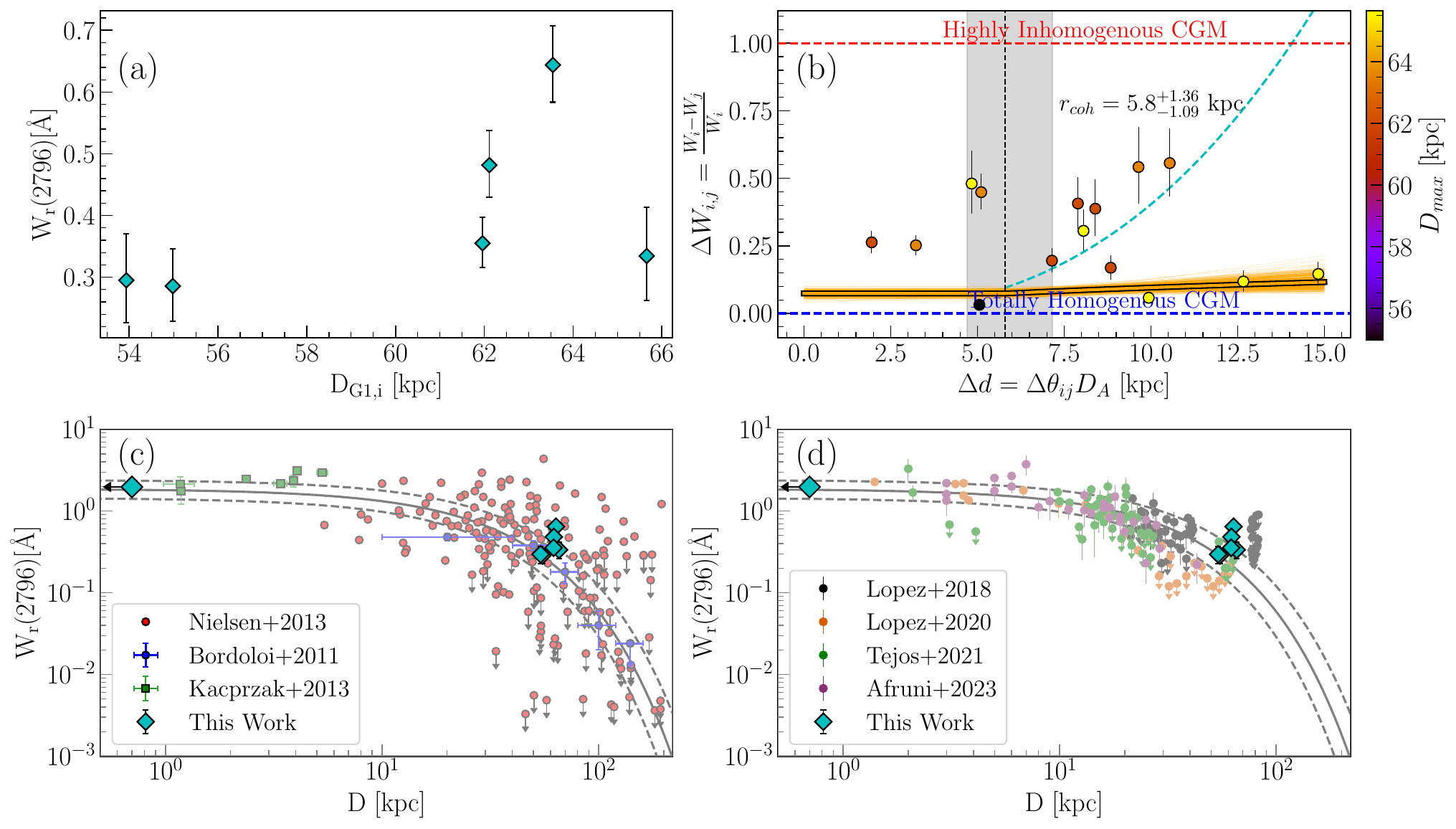}
  \caption{\textit{Panel (a):} Variation in {\mgii} 2796 rest-frame equivalent width ($W_r$) with impact parameter from galaxy G1, detected in the spectra of the background arc in Figure~\ref{fig:wl_rgb_kcwi_muse_hst}. \textit{Panel (b):} The pairwise fractional difference (equation \ref{eq:FD_MgII2796_KCWI}) in $W_r$ versus pairwise physical separation between individual sightlines. The color bar denotes the maximum impact parameter from  G1 between each sightline pair. The solid orange line is the best-fit fractional difference model (equation~\ref{eq:fd_model}) fit to the data. The orange lines are 300 random model realizations drawn from the model parameter posterior distribution. The best-fit coherence length ($r_{coh}$) and the 16$^{th}$ and 84$^{th}$ percentile uncertainties are shown as the vertical dashed line and the gray band. The dashed cyan line is the best-fit model if the outlier sightline at 66 kpc is excluded (see text). \textit{Panel (c)}: Variation in {\mgii} 2796 absorption with distance from their host galaxies. The measurements from this work (cyan diamonds) probe the CGM of galaxy G1 out to $\sim$ 60 kpc. The blue circles show the {\mgii} radial absorption profile obtained from stacked background galaxy spectra from \citet{bordoloi2011radial} at $0.5 \leq z \leq 0.9$. The green squares are from \citet{Kacprzak2013MgIIDistribution} at $z\sim 0.1$ and red circles are from \citet{Nielsen2013MAGICAT1CGMMgII}  at $0.07 \leq z \leq 1.1$. The solid line shows the best fit \mgii\ radial profile from  \citet{Nielsen2013MAGICATIICGM} with the $1\sigma$ bounds on the fit (dashed lines). \textit{Panel (d)}: We compare our measurements to gravitational-arc tomography (ARCTOMO) data from \citet{lopez2018clumpy} (black circles), \citet{Lopez2020SlicingCGM} (orange circles), \citet{Tejos2021TelltaleCGM} (green circles), and \citet{afruni2023CoherenceLensCGM} (purple circles). The solid and dashed black lines are the same as \textit{Panel (c)}. }
  \label{fig:Wr_FD_MgII_arc}
\end{figure*}

Since each sightline intersects the galaxy's CGM at different impact parameters, we compute the variation in CGM absorption strengths between 54 and 66 kpc of galaxy G1. These measurements are presented in Figure~\ref{fig:Wr_FD_MgII_arc}, panel ($a$). Individual absorption strengths vary by a factor of 2.2 between 0.29 {\AA} to 0.64 {\AA} within 10 kpc of each other. The mean absorption strength is $\langle W_r \rangle = $ 0.4 {\AA} at $\langle D \rangle = $ 60 kpc. In addition, the centroid velocities of {\mgii} from the model fits for the arc sightlines do not follow G1's disk kinematics and are scattered around the zero-systemic velocity (see Appendix \ref{app:rotation_kinematics}). This kinematic information along with the fact that G1 is driving a strong galactic outflow and a spatially extended high metallicity emission halo is detected (Figure \ref{fig:metallicity}), indicates that the {\mgii} absorption in the CGM might have originated as entrained material in the galactic outflow. However, as the CGM gas is too diffuse, no spatially extended {\mgii} emission at these impact parameters is detected. Therefore, we cannot rule out an alternative origin of this CGM {\mgii} absorption as ambient halo gas. 

As sightline-to-sightline variations are in the halo of the same galaxy, we quantify the small-scale variation of {\mgii} absorption in the CGM of a single galaxy. We measure the pairwise fractional difference ($\Delta W$) in rest-frame equivalent width between two sightlines ($i, j$) as:
\begin{equation}\label{eq:FD_MgII2796_KCWI}
  \Delta W_{i,j} = \frac{W_{r,i} - W_{r,j}}{W_{r,i}},
\end{equation}
where $W_{r,i} \geq W_{r,j}$ \citep{ellison2004lensedQSO}. For each pair of sightline, the pairwise physical separation is calculated as $\Delta d = \Delta \theta_{ij} D_A $, where $\Delta \theta_{ij}$ is the pairwise angular separation between sightlines in the source-plane of G1, and $D_A$ is the angular diameter distance at the redshift of G1. Figure~\ref{fig:Wr_FD_MgII_arc}, panel ($b$), shows $\Delta W$ as a function of physical separation between two sightlines. A value of $\Delta W = 1$ implies that two sightlines are 100\% different (i.e., the CGM is highly in-homogeneous), while $\Delta W = 0$ implies a fully homogeneous CGM. We observe $\Delta W \approx 0.25$ at a separation of $< 3.5$ kpc which decreases to 0.15 at 5 kpc. This is followed by an increase of 0.4 in $\Delta W$ at $\approx 8$ kpc. We see similar variations between separations of $\Delta d = 8-15$ kpc. This indicates the existence of unresolved small-scale structure in the CGM of the galaxy G1. Indeed, studies of {\mgii} absorbers using high-resolution spectra suggest individual cloud sizes of 100s of parsecs \citep{Rauch2002SmallScale}. 

All the $\Delta W$ values at different separations are $\lessapprox 0.56$. This suggests that on kpc scales, {\mgii} absorption in this CGM is spatially coherent. 

We quantify the large-scale spatial coherence of {\mgii} absorption as follows. The {\mgii} fractional difference as a function of $\Delta d$ can be characterized as:
\begin{equation}
f(\Delta d) =
\begin{cases}
  \delta & ;\ \text{if } \Delta d < r_{coh} \\
  \delta \left( \frac{\Delta d}{r_{coh}} \right)^{\gamma} & ;\ \text{if } \Delta d \geq r_{coh}
\end{cases}
\label{eq:fd_model}
\end{equation}
where $\delta$ is the amplitude of $\Delta W$ at $\Delta d \leq r_{coh}$, and $r_{coh}$ is the coherence length scale, and $\gamma$ is the index of the power-law for $\Delta d \geq r_{coh}$. 

This model is fitted to the observations presented in Figure~\ref{fig:Wr_FD_MgII_arc}, panel ($b$), which yields an amplitude of $\delta = 0.07_{-0.01}^{+0.01}$, $\gamma = 0.48_{-0.27}^{+0.31}$, and a coherence length scale of $r_{coh} = 5.8_{-1.09}^{+1.36}$ kpc, respectively. The best-fit model is presented as the solid orange line in Figure~\ref{fig:Wr_FD_MgII_arc}, panel (\textit{b}). The fainter and thinner orange lines show the 300 random model realizations from the posterior distribution of the fitted model parameters. This model fit is dominated by the 3 data points with low $\Delta W \leq 0.1$ at pairwise separations of $\Delta d \geq 10$ kpc. These values come primarily from one sightline at $\sim$ 65.7 kpc (panel ($a$)). If we exclude this sightline and fit the model, the new best-fit model yields different $\delta = 0.07_{-0.01}^{+0.01}$ and $\gamma = 2.66_{-0.29}^{+0.23}$ but a very similar coherence length of $r_{coh} = 5.23_{-0.2}^{+0.31}$, consistent with the coherence length obtained while fitting the full dataset. This new fit is shown as the dashed cyan line in Figure~\ref{fig:Wr_FD_MgII_arc}, panel ($b$). Therefore, throughout this paper, we adopt a best fit $r_{coh} = 5.8_{-1.09}^{+1.36}$ kpc. This measured coherence length value is consistent with the complex gaseous structure sizes of $ 1.4\ \mathrm{kpc} \leq l \leq 7.8$ kpc that are reported in \citet{afruni2023CoherenceLensCGM}, measured from lensed arc tomography of foreground systems. Since galaxy G1 is driving strong galactic outflow ($V_{out} = -179\ \mathrm{km\ s^{-1}}$), it is plausible that the background sightlines are tracing a bulk flow of outflowing gas in the CGM. This might explain the large and coherent CGM structure observed.

By combining all measurements presented above, we report the {\mgii} absorption radial profile of a single galaxy G1.  Figure~\ref{fig:Wr_FD_MgII_arc}, panel ($c$), shows the {\mgii} absorption radial profile around G1 as a function of impact parameter ($D$, cyan diamonds). At $D= 1$ kpc the {\mgii} absorption strength from down-the-barrel measurement of G1 is reported. Clearly, the {\mgii} absorption strength decreases from $D= 1$ kpc to $D= 50$ kpc. This radial drop-off is similar to the statistical CGM observations of 182 galaxies at $0.07 \leq z \leq 1.1$ in the MAGIICAT survey (red circles) from \citet{Nielsen2013MAGICAT1CGMMgII} and the low impact parameter observations (green squares) from \citet{Kacprzak2013MgIIDistribution}. The best-fit radial profile from \citet{Nielsen2013MAGICATIICGM} is shown as the solid line with dashed lines marking the $1\sigma$ bounds on the fit. This log-linear relation is quantified as $\mathrm{log_{10}(W_r) = \alpha_1\ D + \alpha_2}$, where $\alpha_1 = -0.015\pm 0.002$ and $\alpha_2 = 0.26 \pm 0.11$. Finally, the CGM measurements of star-forming $0.5 \leq z \leq 0.9$ galaxies (blue circles) from \citet{bordoloi2011radial} are presented. We note that the \citep{bordoloi2011radial} observations used stacked spectra of extended background galaxies to measure the {\mgii} absorption radial profile.

The {\mgii} absorption radial profile around a single galaxy is broadly consistent with the average radial absorption profile around a large sample of galaxies at similar redshifts. However, the scatter observed in the CGM of a single galaxy ($\sigma (W_r)$ = 0.13) is much smaller than that observed around the MAGIICAT galaxies. This suggests that such statistical samples have additional degrees of freedom (e.g., stellar/halo mass, star-formation rate, azimuthal angle) that contribute to the large scatter in them \citep{Fernandez-Figueroa2022Orientation, afruni2023CoherenceLensCGM}. Studying the CGM of individual galaxies can reduce many of these uncertainties, allowing a more controlled experiment. This work highlights the power of gravitational lensing in probing the faint CGM of foreground galaxies in a spatially resolved manner at multiple sightlines compared to the pencil beam background quasar sightlines that provide just one measurement per sightline per galaxy.

If the observed CGM of galaxy G1 originates from entrained gas carried out to the CGM by a strong outflow, we can calculate an approximate timescale for this event. Assuming a constant velocity ($V_{out} = 179$ \kms) for the outflowing gas, it requires $\approx 328$ Myr for the outflowing gas to reach an impact parameter of $D \sim 60$ kpc.

Recently, it has become increasingly feasible to use strong gravitational lensing to probe the CGM of individual galaxies. \citep{Lopez2020SlicingCGM,Tejos2021TelltaleCGM, Fernandez-Figueroa2022Orientation, Bordoloi2022}.

\citet{rubin2018AndromedaParachute} report a coherence length scale of $\sim 8-22$ kpc from the {\mgii} absorption in 8 foreground absorbing systems probed by a background quadruply lensed quasar at $z= 2.38$ with KCWI. \citet{augustin2021ClumpinessCGM} used a MUSE observation of a quadruply lensed quasar to probe two absorbing {\mgii} systems at $z \approx 1.2$ and $z\approx 1.4$ and measured coherence length scales of $\sim$ 5-6 kpc possibly corresponding to inflowing filaments in the first system, and $\leq$ 2 kpc corresponding to outflowing clumpy structures in the second system. \citet{Dutta2024LensedQSO} calculate a coherence length scale of 10 kpc in the CGM of foreground individual galaxies, groups (more than 1 galaxy according to their definition), and absorbing systems with $z\sim 0.5 -2$ using two fields background lensed quasars at $z \approx 2.2$ and 2.8 with MUSE observations. Furthermore, they report that high-ionization C IV gas is more coherent than the low-ionization {\mgii} gas. This is an indication of the clumpy nature of the {\mgii} gas. In addition, \citet{afruni2023CoherenceLensCGM} showed that C IV is clumpy on kpc scales, and reported a coherence length scale in the range of 1.4 -- 7.8 kpc using the arc tomography technique \citep{lopez2018clumpy} from a sample of gravitationally lensed galaxies probing absorbing systems at $z\approx 1$. In panel \textit{d}, Figure~\ref{fig:Wr_FD_MgII_arc}, we compare {\mgii} measurements of G1 with the measurements of {\mgii} absorption using the gravitational-arc tomography (ARCTOMO; PI: S. Lopez) from \citet{lopez2018clumpy}, \citet{Lopez2020SlicingCGM}, \citet{Tejos2021TelltaleCGM}, and \citet{afruni2023CoherenceLensCGM}. G1 {\mgii} measurements show a consistent trend with the ARCTOMO data.

One should note that the background sightlines used in this work are spatially extended (unlike pencil beam quasar sightlines). Spatially extended background sightlines in themselves will only have spatially averaged absorption measurements which will smear out information regarding the small-scale gas structure and covering fraction of the gas \citep[see e.g.,][ for discussions on this topic]{bordoloi2011radial,bordoloi2014modeling}. Therefore, it becomes inherently difficult to study small-scale cloud properties using spatially extended background sources. Further, as $r_{coh}$ may vary between different galaxies/CGM; it is advisable to measure $r_{coh}$ independently in individual CGM of galaxies. This might be reflected in the wide range of gas coherence lengths reported in the literature. In future work, this technique will be extended to a large sample of spatially resolved CGM observations, which will enable the study of gas coherence lengths in different galaxy types and environments.

\section{Conclusions} \label{sec:conclusions}

This work uses observations from Keck/KCWI and VLT/MUSE wide field IFU spectrographs in the field of view of a bright gravitationally lensed galaxy at $z\approx 2.76$. The field has a galaxy at $z\approx 0.8$, G1, and {\mgii} absorption is detected in its CGM using the gravitationally lensed galaxy as a background source. This work analyzes both the down-the-barrel spectroscopy of the foreground galaxy and the spectra of the background source. The main conclusions of this work are as follows:

\begin{itemize}
  \item The foreground $z\approx 0.8$ galaxy G1 hosts an extended emission gas halo beyond the stellar disk of the galaxy, traced by {\oii}, {\oiii}, and H$\beta$ emission lines. This halo is spatially extended with a maximum diameter of 33.4 kpc in the plane of G1 at $z = 0.8$ as measured from the {\oii} emission (Figure~\ref{fig:ems_maps_muse}).
  
  \item The {\oii}, {\oiii}, and H$\beta$ emission maps show a velocity gradient consistent with the kinematic signature of a co-rotating gas halo (Figure~\ref{fig:velocity_moments}).

  \item From the emission lines {\oii}, {\oiii}, and H$\beta$ fluxes, we measure the metallicity of the extended emission gas halo to be $\mathrm{12+log(O/H) = 8.99 \pm 0.04}$, which is $3_{-1}^{+1}\%$ greater than the photospheric solar value (Figure~\ref{fig:metallicity}).

  \item G1 is driving a strong galactic outflow traced by blueshifted {\mgii} and {\feii} absorption lines. The mean outflow velocity is measured to be $\approx -179 \ \mathrm{km\ s^{-1}}$ and $\approx - 122 \ \mathrm{km\ s^{-1}}$ from {\mgii} and {\feii} absorption lines, respectively (Figures \ref{fig:MgII_KCWI_MUSE_spec_Model} and \ref{fig:feii_model_spec_kcwi_muse}).

  \item The mass outflow rate for G1 at radius of $R = 5$ kpc is $\dot{M}_{out} \geq 64_{-27}^{+31} \ \mathrm{M_{\odot}\ yr^{-1}}$ as measured from the optically thin {\feii} absorption lines (Figure~\ref{fig:MgII_KCWI_MUSE_spec_Model}), corresponding to a mass loading factor of $\eta \geq 2.1_{-0.9}^{+1.0}$.
  
  \item Using the lensed arc of background galaxy at $z = 2.76$, we detect the {\mgii} absorption doublet in the CGM up to distances of 54--66 kpc away from the center G1. The CGM gas kinematics is inconsistent with an extended disk. The host galaxy is driving a strong galactic outflow and has a high metallicity spatially extended halo (detected in emission). These observations suggest that the {\mgii} absorption detected in the CGM, could have originated as cool gas entrained in galactic outflow from G1. However, an alternative origin of cool ambient halo gas cannot be ruled out. (Figures \ref{fig:wl_rgb_kcwi_muse_hst} and \ref{fig:cgm_spec_kcwi_muse}).

  \item We see small scale variations in the rest-frame equivalent width of the {\mgii} ionized gas in the CGM of G1 at impact parameters around $\approx 60$ kpc, within a region of size $\leq 15$ kpc (Figures \ref{fig:cgm_spec_kcwi_muse} and \ref{fig:Wr_FD_MgII_arc}) 
  
  \item By using a power-law model to fit the fractional difference measurements, we obtain a coherence length scale of the CGM of $ 5.8_{-1.09}^{+1.36}$ kpc (\S \ref{sec:CGM_Absorption_Results} and Figure~\ref{fig:Wr_FD_MgII_arc}).

\end{itemize}

In conclusion, a galaxy at $z\approx 0.8$ is driving a strong galactic outflow, which has a high metallicity co-rotating gas halo extending out to a maximum radius of 16.7 kpc in the plane of G1. The CGM of this galaxy G1 shows {\mgii} absorption that extends out to an impact parameter of 66 kpc in the plane of G1. The CGM absorption is spatially coherent at large scales, with a coherence length of 5.8 kpc. These observations taken together suggest that we may be observing a galaxy, where the galactic outflow is getting recycled, and enriched recycled outflow might be co-rotating around the galaxy before forming the next generation of stars in the galaxy. This scenario is consistent with what is expected in galaxy formation simulations.

This work shows how gravitational lensing, as a cosmic telescope, can be coupled with IFU instruments to probe the small-scale variations in the faint CGM of foreground galaxies. The measurements of such variations will provide strong constraints for galaxy formation and evolution models.

\section{Acknowledgments}
This material is based upon work supported by the National Science Foundation under grant No. NSF AST-2206853. Some data presented herein were obtained at Keck Observatory, which is a private 501(c)3 nonprofit organization operated as a scientific partnership among the California Institute of Technology, the University of California, and the National Aeronautics and Space Administration. The Observatory was made possible by the generous financial support of the W. M. Keck Foundation. The authors wish to recognize and acknowledge the very significant cultural role and reverence that the summit of Maunakea has always had within the Native Hawaiian community. We are most fortunate to have the opportunity to conduct observations from this mountain. This work is based on observations collected at the European Organization for Astronomical Research in the Southern Hemisphere under ESO program 0103.A-0485(B). This research made use of Montage. It is funded by the National Science Foundation under grant No. ACI-1440620, and was previously funded by the National Aeronautics and Space Administration's Earth Science Technology Office, Computation Technologies Project, under Cooperative Agreement Number NCC5-626 between NASA and the California Institute of Technology. Furthermore, we used observations made with the NASA/ESA Hubble Space Telescope, obtained from the data archive at the Space Telescope Science Institute (STScI). STScI is operated by the Association of Universities for Research in Astronomy, Inc. under NASA contract NAS 5-26555. N.T. and S.L. acknowledge support by FONDECYT grant 1231187. All the {\it HST} data used in this paper can be found in MAST: \dataset[10.17909/jgas-n891]{http://dx.doi.org/10.17909/jgas-n891}.

\vspace{5mm}
\facilities{HST(WFC3), Keck: II (KCWI), VLT (MUSE).}

\software{astropy \citep{astropy2013astropy, astropy2018astropy, astropy2022III}, photutils \citep{bradley2023photutils}}

\appendix
\section{CGM gas is not co-rotating with the stellar disk of G1}\label{app:rotation_kinematics}
\begin{figure}
  \centering
  \includegraphics[width=\linewidth]{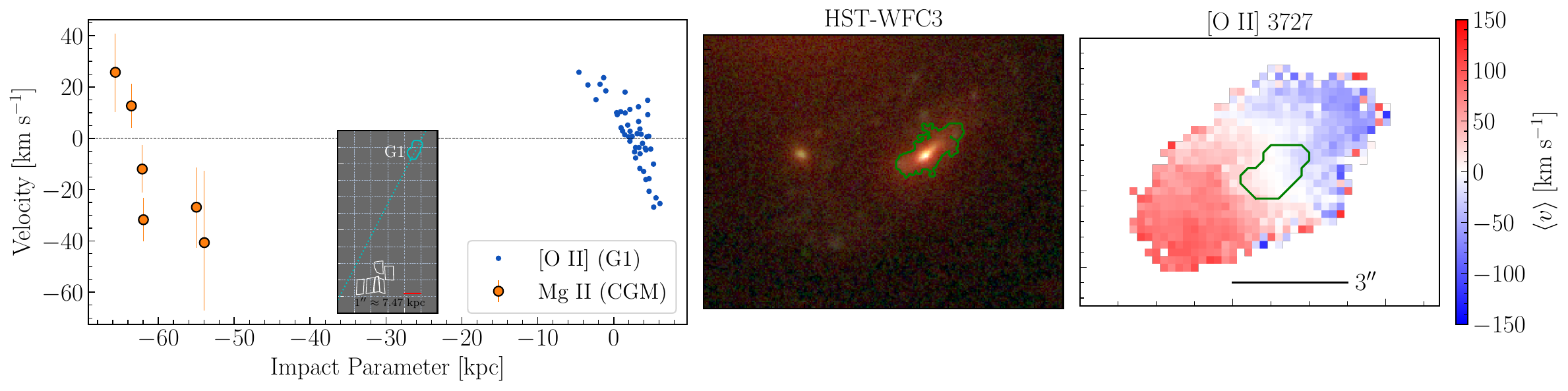}
  \caption{\textit{Left:} G1 kinematics as traced by the {\oii} emission lines (blue circles). The orange data points represent the velocity centroids of {\mgii} absorption lines with respect to the galaxy systemic redshift. The inset shows the ray-traced positions of G1 and the background lines of sight in the absorber plane of G1. The dotted cyan line in the inset marks the major axis of G1, extended out to 60 kpc. \textit{Middle:} HST RGB image of the galaxy G1, with the green contour marking the 3$\sigma$ isophotes of stellar light of the galaxy from the WFC3-UVIS F606W filter. \textit{Right}: 2D velocity map of the {\oii} emission from Figure~\ref{fig:velocity_moments}. The green contour represents the location of the stellar light of the galaxy from the HST image in the middle panel. This Figure qualitatively shows that the {\mgii} absorption in the CGM of G1 is not co-rotating with the stellar disk. The $3^{\prime\prime}$ black line marker corresponds to $\approx$ 10 kpc in the plane of G1.}
  \label{fig:rotation_oii_mgii}
\end{figure}

In this appendix, we calculate the impact parameter of each spaxel with {\oii} emission within the stellar disk of G1 (green contour in Figure~\ref{fig:rotation_oii_mgii}. Any pixel with an x-position to the left of the central pixel of G1 is assigned a negative impact parameter. Figure~\ref{fig:rotation_oii_mgii} shows the kinematics of the stellar disk of G1 as traced by the {\oii} emission lines (blue circles). In addition, we plot the centroid velocities of {\mgii} absorption for the six apertures across the arc. The {\mgii} velocities show both positive and negative values with respect to the systemic redshift of G1. Particularly, the closest apertures show velocities that are not aligned with the kinematics of the stellar disk of the galaxy. This suggests that the {\mgii} absorption is not co-rotating with the disk of G1 (green contour in Figure~\ref{fig:rotation_oii_mgii}). This further supports that the origin of this gas is due to galactic outflow from G1. 

\bibliography{myBibliography}{}

\begin{thebibliography}{}
\expandafter\ifx\csname natexlab\endcsname\relax\def\natexlab#1{#1}\fi
\providecommand{\url}[1]{\href{#1}{#1}}
\providecommand{\dodoi}[1]{doi:~\href{http://doi.org/#1}{\nolinkurl{#1}}}
\providecommand{\doeprint}[1]{\href{http://ascl.net/#1}{\nolinkurl{http://ascl.net/#1}}}
\providecommand{\doarXiv}[1]{\href{https://arxiv.org/abs/#1}{\nolinkurl{https://arxiv.org/abs/#1}}}

\bibitem[{{Afruni} {et~al.}(2023){Afruni}, {Lopez}, {Anshul}, {Tejos}, {Noterdaeme}, {Berg}, {Ledoux}, {Solimano}, {Gonzalez-Lopez}, {Gronke}, {Barrientos}, \& {Johnston}}]{afruni2023CoherenceLensCGM}
{Afruni}, A., {Lopez}, S., {Anshul}, P., {et~al.} 2023, \aap, 680, A112, \dodoi{10.1051/0004-6361/202347867}

\bibitem[{{Asplund} {et~al.}(2021){Asplund}, {Amarsi}, \& {Grevesse}}]{Asplund2021SunComposition}
{Asplund}, M., {Amarsi}, A.~M., \& {Grevesse}, N. 2021, \aap, 653, A141, \dodoi{10.1051/0004-6361/202140445}

\bibitem[{{Astropy Collaboration} {et~al.}(2013){Astropy Collaboration}, {Robitaille}, {Tollerud}, {Greenfield}, {Droettboom}, {Bray}, {Aldcroft}, {Davis}, {Ginsburg}, {Price-Whelan}, {Kerzendorf}, {Conley}, {Crighton}, {Barbary}, {Muna}, {Ferguson}, {Grollier}, {Parikh}, {Nair}, {Unther}, {Deil}, {Woillez}, {Conseil}, {Kramer}, {Turner}, {Singer}, {Fox}, {Weaver}, {Zabalza}, {Edwards}, {Azalee Bostroem}, {Burke}, {Casey}, {Crawford}, {Dencheva}, {Ely}, {Jenness}, {Labrie}, {Lim}, {Pierfederici}, {Pontzen}, {Ptak}, {Refsdal}, {Servillat}, \& {Streicher}}]{astropy2013astropy}
{Astropy Collaboration}, {Robitaille}, T.~P., {Tollerud}, E.~J., {et~al.} 2013, \aap, 558, A33, \dodoi{10.1051/0004-6361/201322068}

\bibitem[{{Astropy Collaboration} {et~al.}(2018){Astropy Collaboration}, {Price-Whelan}, {Sip{\H{o}}cz}, {G{\"u}nther}, {Lim}, {Crawford}, {Conseil}, {Shupe}, {Craig}, {Dencheva}, {Ginsburg}, {VanderPlas}, {Bradley}, {P{\'e}rez-Su{\'a}rez}, {de Val-Borro}, {Aldcroft}, {Cruz}, {Robitaille}, {Tollerud}, {Ardelean}, {Babej}, {Bach}, {Bachetti}, {Bakanov}, {Bamford}, {Barentsen}, {Barmby}, {Baumbach}, {Berry}, {Biscani}, {Boquien}, {Bostroem}, {Bouma}, {Brammer}, {Bray}, {Breytenbach}, {Buddelmeijer}, {Burke}, {Calderone}, {Cano Rodr{\'\i}guez}, {Cara}, {Cardoso}, {Cheedella}, {Copin}, {Corrales}, {Crichton}, {D'Avella}, {Deil}, {Depagne}, {Dietrich}, {Donath}, {Droettboom}, {Earl}, {Erben}, {Fabbro}, {Ferreira}, {Finethy}, {Fox}, {Garrison}, {Gibbons}, {Goldstein}, {Gommers}, {Greco}, {Greenfield}, {Groener}, {Grollier}, {Hagen}, {Hirst}, {Homeier}, {Horton}, {Hosseinzadeh}, {Hu}, {Hunkeler}, {Ivezi{\'c}}, {Jain}, {Jenness}, {Kanarek}, {Kendrew}, {Kern}, {Kerzendorf}, {Khvalko}, {King}, {Kirkby}, {Kulkarni},
  {Kumar}, {Lee}, {Lenz}, {Littlefair}, {Ma}, {Macleod}, {Mastropietro}, {McCully}, {Montagnac}, {Morris}, {Mueller}, {Mumford}, {Muna}, {Murphy}, {Nelson}, {Nguyen}, {Ninan}, {N{\"o}the}, {Ogaz}, {Oh}, {Parejko}, {Parley}, {Pascual}, {Patil}, {Patil}, {Plunkett}, {Prochaska}, {Rastogi}, {Reddy Janga}, {Sabater}, {Sakurikar}, {Seifert}, {Sherbert}, {Sherwood-Taylor}, {Shih}, {Sick}, {Silbiger}, {Singanamalla}, {Singer}, {Sladen}, {Sooley}, {Sornarajah}, {Streicher}, {Teuben}, {Thomas}, {Tremblay}, {Turner}, {Terr{\'o}n}, {van Kerkwijk}, {de la Vega}, {Watkins}, {Weaver}, {Whitmore}, {Woillez}, {Zabalza}, \& {Astropy Contributors}}]{astropy2018astropy}
{Astropy Collaboration}, {Price-Whelan}, A.~M., {Sip{\H{o}}cz}, B.~M., {et~al.} 2018, AJ, 156, 123, \dodoi{10.3847/1538-3881/aabc4f}

\bibitem[{{Astropy Collaboration} {et~al.}(2022){Astropy Collaboration}, {Price-Whelan}, {Lim}, {Earl}, {Starkman}, {Bradley}, {Shupe}, {Patil}, {Corrales}, {Brasseur}, {N{\"o}the}, {Donath}, {Tollerud}, {Morris}, {Ginsburg}, {Vaher}, {Weaver}, {Tocknell}, {Jamieson}, {van Kerkwijk}, {Robitaille}, {Merry}, {Bachetti}, {G{\"u}nther}, {Aldcroft}, {Alvarado-Montes}, {Archibald}, {B{\'o}di}, {Bapat}, {Barentsen}, {Baz{\'a}n}, {Biswas}, {Boquien}, {Burke}, {Cara}, {Cara}, {Conroy}, {Conseil}, {Craig}, {Cross}, {Cruz}, {D'Eugenio}, {Dencheva}, {Devillepoix}, {Dietrich}, {Eigenbrot}, {Erben}, {Ferreira}, {Foreman-Mackey}, {Fox}, {Freij}, {Garg}, {Geda}, {Glattly}, {Gondhalekar}, {Gordon}, {Grant}, {Greenfield}, {Groener}, {Guest}, {Gurovich}, {Handberg}, {Hart}, {Hatfield-Dodds}, {Homeier}, {Hosseinzadeh}, {Jenness}, {Jones}, {Joseph}, {Kalmbach}, {Karamehmetoglu}, {Ka{\l}uszy{\'n}ski}, {Kelley}, {Kern}, {Kerzendorf}, {Koch}, {Kulumani}, {Lee}, {Ly}, {Ma}, {MacBride}, {Maljaars}, {Muna}, {Murphy}, {Norman},
  {O'Steen}, {Oman}, {Pacifici}, {Pascual}, {Pascual-Granado}, {Patil}, {Perren}, {Pickering}, {Rastogi}, {Roulston}, {Ryan}, {Rykoff}, {Sabater}, {Sakurikar}, {Salgado}, {Sanghi}, {Saunders}, {Savchenko}, {Schwardt}, {Seifert-Eckert}, {Shih}, {Jain}, {Shukla}, {Sick}, {Simpson}, {Singanamalla}, {Singer}, {Singhal}, {Sinha}, {Sip{\H{o}}cz}, {Spitler}, {Stansby}, {Streicher}, {{\v{S}}umak}, {Swinbank}, {Taranu}, {Tewary}, {Tremblay}, {de Val-Borro}, {Van Kooten}, {Vasovi{\'c}}, {Verma}, {de Miranda Cardoso}, {Williams}, {Wilson}, {Winkel}, {Wood-Vasey}, {Xue}, {Yoachim}, {Zhang}, {Zonca}, \& {Astropy Project Contributors}}]{astropy2022III}
{Astropy Collaboration}, {Price-Whelan}, A.~M., {Lim}, P.~L., {et~al.} 2022, \apj, 935, 167, \dodoi{10.3847/1538-4357/ac7c74}

\bibitem[{{Augustin} {et~al.}(2021){Augustin}, {P{\'e}roux}, {Hamanowicz}, {Kulkarni}, {Rahmani}, \& {Zanella}}]{augustin2021ClumpinessCGM}
{Augustin}, R., {P{\'e}roux}, C., {Hamanowicz}, A., {et~al.} 2021, \mnras, 505, 6195, \dodoi{10.1093/mnras/stab1673}

\bibitem[{{Bacon} {et~al.}(2010){Bacon}, {Accardo}, {Adjali}, {Anwand}, {Bauer}, {Biswas}, {Blaizot}, {Boudon}, {Brau-Nogue}, {Brinchmann}, {Caillier}, {Capoani}, {Carollo}, {Contini}, {Couderc}, {Daguis{\'e}}, {Deiries}, {Delabre}, {Dreizler}, {Dubois}, {Dupieux}, {Dupuy}, {Emsellem}, {Fechner}, {Fleischmann}, {Fran{\c{c}}ois}, {Gallou}, {Gharsa}, {Glindemann}, {Gojak}, {Guiderdoni}, {Hansali}, {Hahn}, {Jarno}, {Kelz}, {Koehler}, {Kosmalski}, {Laurent}, {Le Floch}, {Lilly}, {Lizon}, {Loupias}, {Manescau}, {Monstein}, {Nicklas}, {Olaya}, {Pares}, {Pasquini}, {P{\'e}contal-Rousset}, {Pell{\'o}}, {Petit}, {Popow}, {Reiss}, {Remillieux}, {Renault}, {Roth}, {Rupprecht}, {Serre}, {Schaye}, {Soucail}, {Steinmetz}, {Streicher}, {Stuik}, {Valentin}, {Vernet}, {Weilbacher}, {Wisotzki}, \& {Yerle}}]{bacon2010muse}
{Bacon}, R., {Accardo}, M., {Adjali}, L., {et~al.} 2010, in Society of Photo-Optical Instrumentation Engineers (SPIE) Conference Series, Vol. 7735, Ground-based and Airborne Instrumentation for Astronomy III, ed. I.~S. {McLean}, S.~K. {Ramsay}, \& H.~{Takami}, 773508, \dodoi{10.1117/12.856027}

\bibitem[{{Bayliss} {et~al.}(2011){Bayliss}, {Hennawi}, {Gladders}, {Koester}, {Sharon}, {Dahle}, \& {Oguri}}]{bayliss2011GeminiLenses}
{Bayliss}, M.~B., {Hennawi}, J.~F., {Gladders}, M.~D., {et~al.} 2011, \apjs, 193, 8, \dodoi{10.1088/0067-0049/193/1/8}

\bibitem[{{Behroozi} {et~al.}(2019){Behroozi}, {Wechsler}, {Hearin}, \& {Conroy}}]{Behroozi2019UniverseMachine}
{Behroozi}, P., {Wechsler}, R.~H., {Hearin}, A.~P., \& {Conroy}, C. 2019, \mnras, 488, 3143, \dodoi{10.1093/mnras/stz1182}

\bibitem[{{Berg} {et~al.}(2024){Berg}, {Afruni}, {Ledoux}, {Lopez}, {Noterdaeme}, {Tejos}, {Hernandez}, {Barrientos}, \& {Johnston}}]{Berg2024HIGravitionalArcs}
{Berg}, T. A.~M., {Afruni}, A., {Ledoux}, C., {et~al.} 2024, arXiv e-prints, arXiv:2412.07652, \dodoi{10.48550/arXiv.2412.07652}

\bibitem[{{Bielby} {et~al.}(2019){Bielby}, {Stott}, {Cullen}, {Tripp}, {Burchett}, {Fumagalli}, {Morris}, {Tejos}, {Crain}, {Bower}, \& {Prochaska}}]{Bielby2019QSAGE}
{Bielby}, R.~M., {Stott}, J.~P., {Cullen}, F., {et~al.} 2019, \mnras, 486, 21, \dodoi{10.1093/mnras/stz774}

\bibitem[{{Bordoloi} {et~al.}(2014{\natexlab{a}}){Bordoloi}, {Lilly}, {Kacprzak}, \& {Churchill}}]{bordoloi2014modeling}
{Bordoloi}, R., {Lilly}, S.~J., {Kacprzak}, G.~G., \& {Churchill}, C.~W. 2014{\natexlab{a}}, \apj, 784, 108, \dodoi{10.1088/0004-637X/784/2/108}

\bibitem[{{Bordoloi} {et~al.}(2016){Bordoloi}, {Rigby}, {Tumlinson}, {Bayliss}, {Sharon}, {Gladders}, \& {Wuyts}}]{bordoloi2016spatially}
{Bordoloi}, R., {Rigby}, J.~R., {Tumlinson}, J., {et~al.} 2016, \mnras, 458, 1891, \dodoi{10.1093/mnras/stw449}

\bibitem[{{Bordoloi} {et~al.}(2011){Bordoloi}, {Lilly}, {Knobel}, {Bolzonella}, {Kampczyk}, {Carollo}, {Iovino}, {Zucca}, {Contini}, {Kneib}, {Le Fevre}, {Mainieri}, {Renzini}, {Scodeggio}, {Zamorani}, {Balestra}, {Bardelli}, {Bongiorno}, {Caputi}, {Cucciati}, {de la Torre}, {de Ravel}, {Garilli}, {Kova{\v{c}}}, {Lamareille}, {Le Borgne}, {Le Brun}, {Maier}, {Mignoli}, {Pello}, {Peng}, {Perez Montero}, {Presotto}, {Scarlata}, {Silverman}, {Tanaka}, {Tasca}, {Tresse}, {Vergani}, {Barnes}, {Cappi}, {Cimatti}, {Coppa}, {Diener}, {Franzetti}, {Koekemoer}, {L{\'o}pez-Sanjuan}, {McCracken}, {Moresco}, {Nair}, {Oesch}, {Pozzetti}, \& {Welikala}}]{bordoloi2011radial}
{Bordoloi}, R., {Lilly}, S.~J., {Knobel}, C., {et~al.} 2011, \apj, 743, 10, \dodoi{10.1088/0004-637X/743/1/10}

\bibitem[{{Bordoloi} {et~al.}(2014{\natexlab{b}}){Bordoloi}, {Lilly}, {Hardmeier}, {Contini}, {Kneib}, {Le Fevre}, {Mainieri}, {Renzini}, {Scodeggio}, {Zamorani}, {Bardelli}, {Bolzonella}, {Bongiorno}, {Caputi}, {Carollo}, {Cucciati}, {de la Torre}, {de Ravel}, {Garilli}, {Iovino}, {Kampczyk}, {Kova{\v{c}}}, {Knobel}, {Lamareille}, {Le Borgne}, {Le Brun}, {Maier}, {Mignoli}, {Oesch}, {Pello}, {Peng}, {Perez Montero}, {Presotto}, {Silverman}, {Tanaka}, {Tasca}, {Tresse}, {Vergani}, {Zucca}, {Cappi}, {Cimatti}, {Coppa}, {Franzetti}, {Koekemoer}, {Moresco}, {Nair}, \& {Pozzetti}}]{bordoloi2014dependence}
{Bordoloi}, R., {Lilly}, S.~J., {Hardmeier}, E., {et~al.} 2014{\natexlab{b}}, \apj, 794, 130, \dodoi{10.1088/0004-637X/794/2/130}

\bibitem[{{Bordoloi} {et~al.}(2022){Bordoloi}, {O'Meara}, {Sharon}, {Rigby}, {Cooke}, {Shaban}, {Matuszewski}, {Rizzi}, {Doppmann}, {Martin}, {Moore}, {Morrissey}, \& {Neill}}]{Bordoloi2022}
{Bordoloi}, R., {O'Meara}, J.~M., {Sharon}, K., {et~al.} 2022, \nat, 606, 59, \dodoi{10.1038/s41586-022-04616-1}

\bibitem[{{Bordoloi} {et~al.}(2024){Bordoloi}, {Simcoe}, {Matthee}, {Kashino}, {Mackenzie}, {Lilly}, {Eilers}, {Liu}, {DePalma}, {Yue}, \& {P. Naidu}}]{Bordoloi2024EIGER4CGM}
{Bordoloi}, R., {Simcoe}, R.~A., {Matthee}, J., {et~al.} 2024, \apj, 963, 28, \dodoi{10.3847/1538-4357/ad1b63}

\bibitem[{Bradley {et~al.}(2023)Bradley, Sip{\H o}cz, Robitaille, Tollerud, Vin{\'{\i}}cius, Deil, Barbary, Wilson, Busko, Donath, G{\"u}nther, Cara, Lim, Me{\ss}linger, Conseil, Bostroem, Droettboom, Bray, Bratholm, Barentsen, Craig, Rathi, Pascual, Perren, Georgiev, de~Val-Borro, Kerzendorf, Bach, Quint, \& Souchereau}]{bradley2023photutils}
Bradley, L., Sip{\H o}cz, B., Robitaille, T., {et~al.} 2023, astropy/photutils: 1.8.0, 1.8.0,  Zenodo, \dodoi{10.5281/zenodo.7946442}

\bibitem[{{Burchett} {et~al.}(2021){Burchett}, {Rubin}, {Prochaska}, {Coil}, {Vaught}, \& {Hennawi}}]{burchett2020circumgalactic}
{Burchett}, J.~N., {Rubin}, K. H.~R., {Prochaska}, J.~X., {et~al.} 2021, \apj, 909, 151, \dodoi{10.3847/1538-4357/abd4e0}

\bibitem[{{Calzetti} {et~al.}(2000){Calzetti}, {Armus}, {Bohlin}, {Kinney}, {Koornneef}, \& {Storchi-Bergmann}}]{Calzetti2000AttenuationLaw}
{Calzetti}, D., {Armus}, L., {Bohlin}, R.~C., {et~al.} 2000, \apj, 533, 682, \dodoi{10.1086/308692}

\bibitem[{{Cardelli} {et~al.}(1989){Cardelli}, {Clayton}, \& {Mathis}}]{Cardelli1989Extinction}
{Cardelli}, J.~A., {Clayton}, G.~C., \& {Mathis}, J.~S. 1989, \apj, 345, 245, \dodoi{10.1086/167900}

\bibitem[{{Chambers} {et~al.}(2016){Chambers}, {Magnier}, {Metcalfe}, {Flewelling}, {Huber}, {Waters}, {Denneau}, {Draper}, {Farrow}, {Finkbeiner}, {Holmberg}, {Koppenhoefer}, {Price}, {Rest}, {Saglia}, {Schlafly}, {Smartt}, {Sweeney}, {Wainscoat}, {Burgett}, {Chastel}, {Grav}, {Heasley}, {Hodapp}, {Jedicke}, {Kaiser}, {Kudritzki}, {Luppino}, {Lupton}, {Monet}, {Morgan}, {Onaka}, {Shiao}, {Stubbs}, {Tonry}, {White}, {Ba{\~n}ados}, {Bell}, {Bender}, {Bernard}, {Boegner}, {Boffi}, {Botticella}, {Calamida}, {Casertano}, {Chen}, {Chen}, {Cole}, {Deacon}, {Frenk}, {Fitzsimmons}, {Gezari}, {Gibbs}, {Goessl}, {Goggia}, {Gourgue}, {Goldman}, {Grant}, {Grebel}, {Hambly}, {Hasinger}, {Heavens}, {Heckman}, {Henderson}, {Henning}, {Holman}, {Hopp}, {Ip}, {Isani}, {Jackson}, {Keyes}, {Koekemoer}, {Kotak}, {Le}, {Liska}, {Long}, {Lucey}, {Liu}, {Martin}, {Masci}, {McLean}, {Mindel}, {Misra}, {Morganson}, {Murphy}, {Obaika}, {Narayan}, {Nieto-Santisteban}, {Norberg}, {Peacock}, {Pier}, {Postman}, {Primak}, {Rae}, {Rai},
  {Riess}, {Riffeser}, {Rix}, {R{\"o}ser}, {Russel}, {Rutz}, {Schilbach}, {Schultz}, {Scolnic}, {Strolger}, {Szalay}, {Seitz}, {Small}, {Smith}, {Soderblom}, {Taylor}, {Thomson}, {Taylor}, {Thakar}, {Thiel}, {Thilker}, {Unger}, {Urata}, {Valenti}, {Wagner}, {Walder}, {Walter}, {Watters}, {Werner}, {Wood-Vasey}, \& {Wyse}}]{Chambers2016PanSTARRS1Surveys}
{Chambers}, K.~C., {Magnier}, E.~A., {Metcalfe}, N., {et~al.} 2016, arXiv e-prints, arXiv:1612.05560, \dodoi{10.48550/arXiv.1612.05560}

\bibitem[{{Chen} {et~al.}(2010){Chen}, {Tremonti}, {Heckman}, {Kauffmann}, {Weiner}, {Brinchmann}, \& {Wang}}]{Chen2010Outflows}
{Chen}, Y.-M., {Tremonti}, C.~A., {Heckman}, T.~M., {et~al.} 2010, \aj, 140, 445, \dodoi{10.1088/0004-6256/140/2/445}

\bibitem[{{Chisholm} {et~al.}(2018){Chisholm}, {Bordoloi}, {Rigby}, \& {Bayliss}}]{chisholm2018feeding}
{Chisholm}, J., {Bordoloi}, R., {Rigby}, J.~R., \& {Bayliss}, M. 2018, \mnras, 474, 1688, \dodoi{10.1093/mnras/stx2848}

\bibitem[{{Chisholm} {et~al.}(2016){Chisholm}, {Tremonti}, {Leitherer}, {Chen}, \& {Wofford}}]{chisholm2016shining}
{Chisholm}, J., {Tremonti}, C.~A., {Leitherer}, C., {Chen}, Y., \& {Wofford}, A. 2016, \mnras, 457, 3133, \dodoi{10.1093/mnras/stw178}

\bibitem[{{Chisholm} {et~al.}(2015){Chisholm}, {Tremonti}, {Leitherer}, {Chen}, {Wofford}, \& {Lundgren}}]{chisholm2015scaling}
{Chisholm}, J., {Tremonti}, C.~A., {Leitherer}, C., {et~al.} 2015, \apj, 811, 149, \dodoi{10.1088/0004-637X/811/2/149}

\bibitem[{{Clark} {et~al.}(2022){Clark}, {Bordoloi}, \& {Fox}}]{Clark2022GasFlowsMW}
{Clark}, S., {Bordoloi}, R., \& {Fox}, A.~J. 2022, \mnras, 512, 811, \dodoi{10.1093/mnras/stac504}

\bibitem[{{Conroy} \& {Gunn}(2010)}]{Conroy2010FSPS}
{Conroy}, C., \& {Gunn}, J.~E. 2010, \apj, 712, 833, \dodoi{10.1088/0004-637X/712/2/833}

\bibitem[{{Conroy} {et~al.}(2009){Conroy}, {Gunn}, \& {White}}]{Conroy2009FSPS}
{Conroy}, C., {Gunn}, J.~E., \& {White}, M. 2009, \apj, 699, 486, \dodoi{10.1088/0004-637X/699/1/486}

\bibitem[{{Doi} {et~al.}(2010){Doi}, {Tanaka}, {Fukugita}, {Gunn}, {Yasuda}, {Ivezi{\'c}}, {Brinkmann}, {de Haars}, {Kleinman}, {Krzesinski}, \& {French Leger}}]{Doi2010SDSSPotoResponse}
{Doi}, M., {Tanaka}, M., {Fukugita}, M., {et~al.} 2010, \aj, 139, 1628, \dodoi{10.1088/0004-6256/139/4/1628}

\bibitem[{{Draine}(2011)}]{Draine2011PhysicsISMIGM}
{Draine}, B.~T. 2011, {Physics of the Interstellar and Intergalactic Medium} (Princeton University Press)

\bibitem[{{Dutta} {et~al.}(2024){Dutta}, {Acebron}, {Fumagalli}, {Grillo}, {Caminha}, \& {Fossati}}]{Dutta2024LensedQSO}
{Dutta}, R., {Acebron}, A., {Fumagalli}, M., {et~al.} 2024, \mnras, \dodoi{10.1093/mnras/stae048}

\bibitem[{{Dutta} {et~al.}(2023){Dutta}, {Fossati}, {Fumagalli}, {Revalski}, {Lofthouse}, {Nelson}, {Papini}, {Rafelski}, {Cantalupo}, {Battaia}, {Dayal}, {Longobardi}, {P{\'e}roux}, {Prichard}, \& {Prochaska}}]{Dutta2023EmissionHalos}
{Dutta}, R., {Fossati}, M., {Fumagalli}, M., {et~al.} 2023, \mnras, \dodoi{10.1093/mnras/stad1002}

\bibitem[{{Ellison} {et~al.}(2004){Ellison}, {Ibata}, {Pettini}, {Lewis}, {Aracil}, {Petitjean}, \& {Srianand}}]{ellison2004lensedQSO}
{Ellison}, S.~L., {Ibata}, R., {Pettini}, M., {et~al.} 2004, \aap, 414, 79, \dodoi{10.1051/0004-6361:20034003}

\bibitem[{{Faucher-Gigu{\`e}re} \& {Oh}(2023)}]{Faucher-Giguere2023CGM}
{Faucher-Gigu{\`e}re}, C.-A., \& {Oh}, S.~P. 2023, \araa, 61, 131, \dodoi{10.1146/annurev-astro-052920-125203}

\bibitem[{{Fernandez-Figueroa} {et~al.}(2022){Fernandez-Figueroa}, {Lopez}, {Tejos}, {Berg}, {Ledoux}, {Noterdaeme}, {Afruni}, {Barrientos}, {Gonzalez-Lopez}, {Hamel}, {Johnston}, {Katsianis}, {Sharon}, \& {Solimano}}]{Fernandez-Figueroa2022Orientation}
{Fernandez-Figueroa}, A., {Lopez}, S., {Tejos}, N., {et~al.} 2022, \mnras, 517, 2214, \dodoi{10.1093/mnras/stac2851}

\bibitem[{{Fitzpatrick}(1999)}]{Fitzpatrick1999Extinction}
{Fitzpatrick}, E.~L. 1999, \pasp, 111, 63, \dodoi{10.1086/316293}

\bibitem[{{Fluetsch} {et~al.}(2021){Fluetsch}, {Maiolino}, {Carniani}, {Arribas}, {Belfiore}, {Bellocchi}, {Cazzoli}, {Cicone}, {Cresci}, {Fabian}, {Gallagher}, {Ishibashi}, {Mannucci}, {Marconi}, {Perna}, {Sturm}, \& {Venturi}}]{Fluetsch2021MultiPhaseOutflows}
{Fluetsch}, A., {Maiolino}, R., {Carniani}, S., {et~al.} 2021, \mnras, 505, 5753, \dodoi{10.1093/mnras/stab1666}

\bibitem[{{Foreman-Mackey} {et~al.}(2013){Foreman-Mackey}, {Hogg}, {Lang}, \& {Goodman}}]{foreman2013emcee}
{Foreman-Mackey}, D., {Hogg}, D.~W., {Lang}, D., \& {Goodman}, J. 2013, \pasp, 125, 306, \dodoi{10.1086/670067}

\bibitem[{{Fox} {et~al.}(2019){Fox}, {Richter}, {Ashley}, {Heckman}, {Lehner}, {Werk}, {Bordoloi}, \& {Peeples}}]{Fox2019MilkyWayOutflows}
{Fox}, A.~J., {Richter}, P., {Ashley}, T., {et~al.} 2019, \apj, 884, 53, \dodoi{10.3847/1538-4357/ab40ad}

\bibitem[{{Fukugita} {et~al.}(1996){Fukugita}, {Ichikawa}, {Gunn}, {Doi}, {Shimasaku}, \& {Schneider}}]{Fukugita1996SDSSPhoto}
{Fukugita}, M., {Ichikawa}, T., {Gunn}, J.~E., {et~al.} 1996, \aj, 111, 1748, \dodoi{10.1086/117915}

\bibitem[{{Hafen} {et~al.}(2019){Hafen}, {Faucher-Gigu{\`e}re}, {Angl{\'e}s-Alc{\'a}zar}, {Stern}, {Kere{\v{s}}}, {Hummels}, {Esmerian}, {Garrison-Kimmel}, {El-Badry}, {Wetzel}, {Chan}, {Hopkins}, \& {Murray}}]{Hafen2019CGMoriginsFIRE}
{Hafen}, Z., {Faucher-Gigu{\`e}re}, C.-A., {Angl{\'e}s-Alc{\'a}zar}, D., {et~al.} 2019, \mnras, 488, 1248, \dodoi{10.1093/mnras/stz1773}

\bibitem[{{Heckman} {et~al.}(2015){Heckman}, {Alexandroff}, {Borthakur}, {Overzier}, \& {Leitherer}}]{heckman2015systematic}
{Heckman}, T.~M., {Alexandroff}, R.~M., {Borthakur}, S., {Overzier}, R., \& {Leitherer}, C. 2015, \apj, 809, 147, \dodoi{10.1088/0004-637X/809/2/147}

\bibitem[{{Heckman} {et~al.}(2000){Heckman}, {Lehnert}, {Strickland}, \& {Armus}}]{Heckman2000GalacticSuperwinds}
{Heckman}, T.~M., {Lehnert}, M.~D., {Strickland}, D.~K., \& {Armus}, L. 2000, \apjs, 129, 493, \dodoi{10.1086/313421}

\bibitem[{{Ho} {et~al.}(2017){Ho}, {Martin}, {Kacprzak}, \& {Churchill}}]{Ho2017GasAccretion}
{Ho}, S.~H., {Martin}, C.~L., {Kacprzak}, G.~G., \& {Churchill}, C.~W. 2017, \apj, 835, 267, \dodoi{10.3847/1538-4357/835/2/267}

\bibitem[{{Hogg}(1999)}]{hogg1999distance}
{Hogg}, D.~W. 1999, arXiv e-prints, astro.
\newblock \doarXiv{astro-ph/9905116}

\bibitem[{{Hopkins} {et~al.}(2014){Hopkins}, {Kere{\v{s}}}, {O{\~n}orbe}, {Faucher-Gigu{\`e}re}, {Quataert}, {Murray}, \& {Bullock}}]{hopkins2014galaxies}
{Hopkins}, P.~F., {Kere{\v{s}}}, D., {O{\~n}orbe}, J., {et~al.} 2014, \mnras, 445, 581, \dodoi{10.1093/mnras/stu1738}

\bibitem[{{Hopkins} {et~al.}(2010){Hopkins}, {Murray}, {Quataert}, \& {Thompson}}]{Hopkins2010MaximumSFDensity}
{Hopkins}, P.~F., {Murray}, N., {Quataert}, E., \& {Thompson}, T.~A. 2010, \mnras, 401, L19, \dodoi{10.1111/j.1745-3933.2009.00777.x}

\bibitem[{{Hopkins} {et~al.}(2012){Hopkins}, {Quataert}, \& {Murray}}]{hopkins2012stellar}
{Hopkins}, P.~F., {Quataert}, E., \& {Murray}, N. 2012, \mnras, 421, 3522, \dodoi{10.1111/j.1365-2966.2012.20593.x}

\bibitem[{{Jacob} {et~al.}(2010{\natexlab{a}}){Jacob}, {Katz}, {Berriman}, {Good}, {Laity}, {Deelman}, {Kesselman}, {Singh}, {Su}, {Prince}, \& {Williams}}]{Jacob2010Montage}
{Jacob}, J.~C., {Katz}, D.~S., {Berriman}, G.~B., {et~al.} 2010{\natexlab{a}}, arXiv e-prints, arXiv:1005.4454, \dodoi{10.48550/arXiv.1005.4454}

\bibitem[{{Jacob} {et~al.}(2010{\natexlab{b}}){Jacob}, {Katz}, {Berriman}, {Good}, {Laity}, {Deelman}, {Kesselman}, {Singh}, {Su}, {Prince}, \& {Williams}}]{Jacob2010MontageCode}
---. 2010{\natexlab{b}}, {Montage: An Astronomical Image Mosaicking Toolkit}, Astrophysics Source Code Library, record ascl:1010.036

\bibitem[{{Jenkins}(2009)}]{jenkins2009unified}
{Jenkins}, E.~B. 2009, \apj, 700, 1299, \dodoi{10.1088/0004-637X/700/2/1299}

\bibitem[{Johnson {et~al.}(2023)Johnson, Foreman-Mackey, Sick, Leja, Walmsley, Tollerud, Leung, Scott, \& Park}]{Johnson2023PyFSPS}
Johnson, B., Foreman-Mackey, D., Sick, J., {et~al.} 2023, dfm/python-fsps: v0.4.6, v0.4.6,  Zenodo, \dodoi{10.5281/zenodo.10026684}

\bibitem[{{Johnson} {et~al.}(2021){Johnson}, {Leja}, {Conroy}, \& {Speagle}}]{johnson2021propspector}
{Johnson}, B.~D., {Leja}, J., {Conroy}, C., \& {Speagle}, J.~S. 2021, \apjs, 254, 22, \dodoi{10.3847/1538-4365/abef67}

\bibitem[{{Kacprzak} {et~al.}(2013){Kacprzak}, {Cooke}, {Churchill}, {Ryan-Weber}, \& {Nielsen}}]{Kacprzak2013MgIIDistribution}
{Kacprzak}, G.~G., {Cooke}, J., {Churchill}, C.~W., {Ryan-Weber}, E.~V., \& {Nielsen}, N.~M. 2013, \apjl, 777, L11, \dodoi{10.1088/2041-8205/777/1/L11}

\bibitem[{{Keerthi Vasan} {et~al.}(2023){Keerthi Vasan}, {Jones}, {Sanders}, {Ellis}, {Stark}, {Kacprzak}, {Barone}, {Tran}, {Glazebrook}, \& {Jacobs}}]{KeerthiVasan2023OutflowsCosmicNoon}
{Keerthi Vasan}, G.~C., {Jones}, T., {Sanders}, R.~L., {et~al.} 2023, \apj, 959, 124, \dodoi{10.3847/1538-4357/acf462}

\bibitem[{{Keerthi Vasan G.} {et~al.}(2024){Keerthi Vasan G.}, {Jones}, {Shajib}, {Rhoades}, {Chen}, {Sanders}, {Stark}, {Ellis}, {Leethochawalit}, {Kacprzak}, {Barone}, {Glazebrook}, {Tran}, {Skobe}, {Mortensen}, \& {Barisic}}]{Vasan2024LensedGalaxy}
{Keerthi Vasan G.}, C., {Jones}, T., {Shajib}, A.~J., {et~al.} 2024, arXiv e-prints, arXiv:2402.00942, \dodoi{10.48550/arXiv.2402.00942}

\bibitem[{{Kennicutt}(1998)}]{kennicutt1998star}
{Kennicutt}, Robert~C., J. 1998, \araa, 36, 189, \dodoi{10.1146/annurev.astro.36.1.189}

\bibitem[{{Kewley} \& {Dopita}(2002)}]{Kewley2002LinesAbundances}
{Kewley}, L.~J., \& {Dopita}, M.~A. 2002, \apjs, 142, 35, \dodoi{10.1086/341326}

\bibitem[{{Koester} {et~al.}(2010){Koester}, {Gladders}, {Hennawi}, {Sharon}, {Wuyts}, {Rigby}, {Bayliss}, \& {Dahle}}]{Koester2010TwoLensed}
{Koester}, B.~P., {Gladders}, M.~D., {Hennawi}, J.~F., {et~al.} 2010, \apjl, 723, L73, \dodoi{10.1088/2041-8205/723/1/L73}

\bibitem[{{Koposov} {et~al.}(2023){Koposov}, {Speagle}, {Barbary}, {Ashton}, {Bennett}, {Buchner}, {Scheffler}, {Cook}, {Talbot}, {Guillochon}, {Cubillos}, {Asensio Ramos}, {Johnson}, {Lang}, {Ilya}, {Dartiailh}, {Nitz}, {McCluskey}, \& {Archibald}}]{Koposov2023dyensty}
{Koposov}, S., {Speagle}, J., {Barbary}, K., {et~al.} 2023, {joshspeagle/dynesty: v2.1.3}, v2.1.3,  Zenodo, \dodoi{10.5281/zenodo.8408702}

\bibitem[{{Kusakabe} {et~al.}(2024){Kusakabe}, {Mauerhofer}, {Verhamme}, {Garel}, {Blaizot}, {Wisotzki}, {Richard}, {Boogaard}, {Leclercq}, {Guo}, {Claeyssens}, {Contini}, {Herenz}, {Kerutt}, {Maseda}, {Michel-Dansac}, {Nanayakkara}, {Ouchi}, {Pessa}, \& {Schaye}}]{Kusakabe2024MXDFCGM}
{Kusakabe}, H., {Mauerhofer}, V., {Verhamme}, A., {et~al.} 2024, arXiv e-prints, arXiv:2406.04399, \dodoi{10.48550/arXiv.2406.04399}

\bibitem[{{Leclercq} {et~al.}(2022){Leclercq}, {Verhamme}, {Epinat}, {Simmonds}, {Matthee}, {Bouch{\'e}}, {Garel}, {Urrutia}, {Wisotzki}, {Zabl}, {Bacon}, {Abril-Melgarejo}, {Boogaard}, {Brinchmann}, {Cantalupo}, {Contini}, {Kerutt}, {Kusakabe}, {Maseda}, {Michel-Dansac}, {Muzahid}, {Nanayakkara}, {Richard}, \& {Schaye}}]{Leclercq2022MUSEMgII_IGrM}
{Leclercq}, F., {Verhamme}, A., {Epinat}, B., {et~al.} 2022, \aap, 663, A11, \dodoi{10.1051/0004-6361/202142179}

\bibitem[{{Lehner} {et~al.}(2022){Lehner}, {Kopenhafer}, {O'Meara}, {Howk}, {Fumagalli}, {Prochaska}, {Acharyya}, {O'Shea}, {Peeples}, {Tumlinson}, \& {Hummels}}]{Lehner2022CGM}
{Lehner}, N., {Kopenhafer}, C., {O'Meara}, J.~M., {et~al.} 2022, \apj, 936, 156, \dodoi{10.3847/1538-4357/ac7400}

\bibitem[{{Leitherer} {et~al.}(2011){Leitherer}, {Tremonti}, {Heckman}, \& {Calzetti}}]{leitherer2011ultraviolet}
{Leitherer}, C., {Tremonti}, C.~A., {Heckman}, T.~M., \& {Calzetti}, D. 2011, \aj, 141, 37, \dodoi{10.1088/0004-6256/141/2/37}

\bibitem[{{Lodders}(2003)}]{lodders2003solar}
{Lodders}, K. 2003, \apj, 591, 1220, \dodoi{10.1086/375492}

\bibitem[{{Lopez} {et~al.}(2018){Lopez}, {Tejos}, {Ledoux}, {Barrientos}, {Sharon}, {Rigby}, {Gladders}, {Bayliss}, \& {Pessa}}]{lopez2018clumpy}
{Lopez}, S., {Tejos}, N., {Ledoux}, C., {et~al.} 2018, \nat, 554, 493, \dodoi{10.1038/nature25436}

\bibitem[{{Lopez} {et~al.}(2020){Lopez}, {Tejos}, {Barrientos}, {Ledoux}, {Sharon}, {Katsianis}, {Florian}, {Rivera-Thorsen}, {Bayliss}, {Dahle}, {Fernandez-Figueroa}, {Gladders}, {Gronke}, {Hamel}, {Pessa}, \& {Rigby}}]{Lopez2020SlicingCGM}
{Lopez}, S., {Tejos}, N., {Barrientos}, L.~F., {et~al.} 2020, \mnras, 491, 4442, \dodoi{10.1093/mnras/stz3183}

\bibitem[{{Lopez} {et~al.}(2024){Lopez}, {Afruni}, {Zamora}, {Tejos}, {Ledoux}, {Hernandez}, {Berg}, {Cortes}, {Urbina}, {Johnston}, {Barrientos}, {Bayliss}, {Cuellar}, {Krogager}, {Noterdaeme}, \& {Solimano}}]{Lopez2024CGMabsorptionCIV}
{Lopez}, S., {Afruni}, A., {Zamora}, D., {et~al.} 2024, \aap, 691, A356, \dodoi{10.1051/0004-6361/202451200}

\bibitem[{{Ma} {et~al.}(2016){Ma}, {Hopkins}, {Faucher-Gigu{\`e}re}, {Zolman}, {Muratov}, {Kere{\v{s}}}, \& {Quataert}}]{Ma2016MZrelation}
{Ma}, X., {Hopkins}, P.~F., {Faucher-Gigu{\`e}re}, C.-A., {et~al.} 2016, \mnras, 456, 2140, \dodoi{10.1093/mnras/stv2659}

\bibitem[{{Martin} {et~al.}(2019){Martin}, {Ho}, {Kacprzak}, \& {Churchill}}]{Martin2019CGMKinematics}
{Martin}, C.~L., {Ho}, S.~H., {Kacprzak}, G.~G., \& {Churchill}, C.~W. 2019, \apj, 878, 84, \dodoi{10.3847/1538-4357/ab18ac}

\bibitem[{{Morrissey} {et~al.}(2018){Morrissey}, {Matuszewski}, {Martin}, {Neill}, {Epps}, {Fucik}, {Weber}, {Darvish}, {Adkins}, {Allen}, {Bartos}, {Belicki}, {Cabak}, {Callahan}, {Cowley}, {Crabill}, {Deich}, {Delecroix}, {Doppman}, {Hilyard}, {James}, {Kaye}, {Kokorowski}, {Kwok}, {Lanclos}, {Milner}, {Moore}, {O'Sullivan}, {Parihar}, {Park}, {Phillips}, {Rizzi}, {Rockosi}, {Rodriguez}, {Salaun}, {Seaman}, {Sheikh}, {Weiss}, \& {Zarzaca}}]{morrissey2018keck}
{Morrissey}, P., {Matuszewski}, M., {Martin}, D.~C., {et~al.} 2018, \apj, 864, 93, \dodoi{10.3847/1538-4357/aad597}

\bibitem[{{Morton}(2003)}]{morton2003atomic}
{Morton}, D.~C. 2003, \apjs, 149, 205, \dodoi{10.1086/377639}

\bibitem[{{Muratov} {et~al.}(2017){Muratov}, {Kere{\v{s}}}, {Faucher-Gigu{\`e}re}, {Hopkins}, {Ma}, {Angl{\'e}s-Alc{\'a}zar}, {Chan}, {Torrey}, {Hafen}, {Quataert}, \& {Murray}}]{muratov2017metalflowsCGM}
{Muratov}, A.~L., {Kere{\v{s}}}, D., {Faucher-Gigu{\`e}re}, C.-A., {et~al.} 2017, \mnras, 468, 4170, \dodoi{10.1093/mnras/stx667}

\bibitem[{{Naab} \& {Ostriker}(2017)}]{Naab2017TheoreticalChallenges}
{Naab}, T., \& {Ostriker}, J.~P. 2017, \araa, 55, 59, \dodoi{10.1146/annurev-astro-081913-040019}

\bibitem[{{Neill} {et~al.}(2023){Neill}, {Matuszewski}, {Martin}, {Brodheim}, \& {Rizzi}}]{Neill2023KCWIDRPPython}
{Neill}, D., {Matuszewski}, M., {Martin}, C., {Brodheim}, M., \& {Rizzi}, L. 2023, {KCWI\_DRP: Keck Cosmic Web Imager Data Reduction Pipeline in Python}, Astrophysics Source Code Library, record ascl:2301.019

\bibitem[{{Nielsen} {et~al.}(2013{\natexlab{a}}){Nielsen}, {Churchill}, \& {Kacprzak}}]{Nielsen2013MAGICATIICGM}
{Nielsen}, N.~M., {Churchill}, C.~W., \& {Kacprzak}, G.~G. 2013{\natexlab{a}}, \apj, 776, 115, \dodoi{10.1088/0004-637X/776/2/115}

\bibitem[{{Nielsen} {et~al.}(2013{\natexlab{b}}){Nielsen}, {Churchill}, {Kacprzak}, \& {Murphy}}]{Nielsen2013MAGICAT1CGMMgII}
{Nielsen}, N.~M., {Churchill}, C.~W., {Kacprzak}, G.~G., \& {Murphy}, M.~T. 2013{\natexlab{b}}, \apj, 776, 114, \dodoi{10.1088/0004-637X/776/2/114}

\bibitem[{O'Meara {et~al.}(2022)O'Meara, Prochaska, \& Bordoloi}]{omeara2022kcwitools}
O'Meara, J., Prochaska, J.~X., \& Bordoloi, R. 2022, pypeit/kcwitools: v0.1.0, v0.1.0,  Zenodo, \dodoi{10.5281/zenodo.6079396}

\bibitem[{{Oppenheimer} \& {Dav{\'e}}(2006)}]{Oppenheimer2006SimulationsOutflows}
{Oppenheimer}, B.~D., \& {Dav{\'e}}, R. 2006, \mnras, 373, 1265, \dodoi{10.1111/j.1365-2966.2006.10989.x}

\bibitem[{{Osterbrock} \& {Ferland}(2006)}]{osterbrock2006astrophysics}
{Osterbrock}, D.~E., \& {Ferland}, G.~J. 2006, {Astrophysics of gaseous nebulae and active galactic nuclei} (University Science Books, Sausalito, CA)

\bibitem[{{Pagel} {et~al.}(1979){Pagel}, {Edmunds}, {Blackwell}, {Chun}, \& {Smith}}]{Pagel1979R23Metallcity}
{Pagel}, B.~E.~J., {Edmunds}, M.~G., {Blackwell}, D.~E., {Chun}, M.~S., \& {Smith}, G. 1979, \mnras, 189, 95, \dodoi{10.1093/mnras/189.1.95}

\bibitem[{{Peeples} {et~al.}(2014){Peeples}, {Werk}, {Tumlinson}, {Oppenheimer}, {Prochaska}, {Katz}, \& {Weinberg}}]{Peeples2014MetalsCGM}
{Peeples}, M.~S., {Werk}, J.~K., {Tumlinson}, J., {et~al.} 2014, \apj, 786, 54, \dodoi{10.1088/0004-637X/786/1/54}

\bibitem[{{P{\'e}roux} \& {Howk}(2020)}]{Peroux2020CosmicBaryon}
{P{\'e}roux}, C., \& {Howk}, J.~C. 2020, \araa, 58, 363, \dodoi{10.1146/annurev-astro-021820-120014}

\bibitem[{{Perrotta} {et~al.}(2024){Perrotta}, {Coil}, {Rupke}, {Ning}, {Duong}, {Diamond-Stanic}, {Fielding}, {Geach}, {Hickox}, {Moustakas}, {Rudnick}, {Sell}, {Swiggum}, \& {Tremonti}}]{Perrota2024OIINebulae}
{Perrotta}, S., {Coil}, A.~L., {Rupke}, D. S.~N., {et~al.} 2024, arXiv e-prints, arXiv:2409.10013, \dodoi{10.48550/arXiv.2409.10013}

\bibitem[{{Pillepich} {et~al.}(2018){Pillepich}, {Springel}, {Nelson}, {Genel}, {Naiman}, {Pakmor}, {Hernquist}, {Torrey}, {Vogelsberger}, {Weinberger}, \& {Marinacci}}]{Pillepich2018SimulatingGFormation}
{Pillepich}, A., {Springel}, V., {Nelson}, D., {et~al.} 2018, \mnras, 473, 4077, \dodoi{10.1093/mnras/stx2656}

\bibitem[{{Rauch} {et~al.}(2002){Rauch}, {Sargent}, {Barlow}, \& {Simcoe}}]{Rauch2002SmallScale}
{Rauch}, M., {Sargent}, W. L.~W., {Barlow}, T.~A., \& {Simcoe}, R.~A. 2002, \apj, 576, 45, \dodoi{10.1086/341267}

\bibitem[{{Reichardt Chu} {et~al.}(2022){Reichardt Chu}, {Fisher}, {Nielsen}, {Chisholm}, {Girard}, {Kacprzak}, {Bolatto}, {Herrera-Camus}, {Sandstrom}, {Li}, {Rickards Vaught}, \& {McPherson}}]{ReichardtChu2022DuvetOutflows}
{Reichardt Chu}, B., {Fisher}, D.~B., {Nielsen}, N.~M., {et~al.} 2022, \mnras, 511, 5782, \dodoi{10.1093/mnras/stac420}

\bibitem[{{Rigby} {et~al.}(2018{\natexlab{a}}){Rigby}, {Bayliss}, {Sharon}, {Gladders}, {Chisholm}, {Dahle}, {Johnson}, {Paterno-Mahler}, {Wuyts}, \& {Kelson}}]{rigby2018magellanI}
{Rigby}, J.~R., {Bayliss}, M.~B., {Sharon}, K., {et~al.} 2018{\natexlab{a}}, \aj, 155, 104, \dodoi{10.3847/1538-3881/aaa2ff}

\bibitem[{{Rigby} {et~al.}(2018{\natexlab{b}}){Rigby}, {Bayliss}, {Chisholm}, {Bordoloi}, {Sharon}, {Gladders}, {Johnson}, {Paterno-Mahler}, {Wuyts}, {Dahle}, \& {Acharyya}}]{rigby2018magellanII}
{Rigby}, J.~R., {Bayliss}, M.~B., {Chisholm}, J., {et~al.} 2018{\natexlab{b}}, \apj, 853, 87, \dodoi{10.3847/1538-4357/aaa2fc}

\bibitem[{{Rubin} {et~al.}(2014){Rubin}, {Prochaska}, {Koo}, {Phillips}, {Martin}, \& {Winstrom}}]{Rubin2014UpiquitousOutflows}
{Rubin}, K. H.~R., {Prochaska}, J.~X., {Koo}, D.~C., {et~al.} 2014, \apj, 794, 156, \dodoi{10.1088/0004-637X/794/2/156}

\bibitem[{{Rubin} {et~al.}(2010{\natexlab{a}}){Rubin}, {Prochaska}, {Koo}, {Phillips}, \& {Weiner}}]{rubin2010galaxies}
{Rubin}, K. H.~R., {Prochaska}, J.~X., {Koo}, D.~C., {Phillips}, A.~C., \& {Weiner}, B.~J. 2010{\natexlab{a}}, \apj, 712, 574, \dodoi{10.1088/0004-637X/712/1/574}

\bibitem[{{Rubin} {et~al.}(2010{\natexlab{b}}){Rubin}, {Weiner}, {Koo}, {Martin}, {Prochaska}, {Coil}, \& {Newman}}]{Rubin2010PersistenceWinds}
{Rubin}, K. H.~R., {Weiner}, B.~J., {Koo}, D.~C., {et~al.} 2010{\natexlab{b}}, \apj, 719, 1503, \dodoi{10.1088/0004-637X/719/2/1503}

\bibitem[{{Rubin} {et~al.}(2018){Rubin}, {O'Meara}, {Cooksey}, {Matuszewski}, {Rizzi}, {Doppmann}, {Kwok}, {Martin}, {Moore}, {Morrissey}, \& {Neill}}]{rubin2018AndromedaParachute}
{Rubin}, K. H.~R., {O'Meara}, J.~M., {Cooksey}, K.~L., {et~al.} 2018, \apj, 859, 146, \dodoi{10.3847/1538-4357/aaaeb7}

\bibitem[{{Rudie} {et~al.}(2012){Rudie}, {Steidel}, {Trainor}, {Rakic}, {Bogosavljevi{\'c}}, {Pettini}, {Reddy}, {Shapley}, {Erb}, \& {Law}}]{Rudie2012CGMHI}
{Rudie}, G.~C., {Steidel}, C.~C., {Trainor}, R.~F., {et~al.} 2012, \apj, 750, 67, \dodoi{10.1088/0004-637X/750/1/67}

\bibitem[{{Rupke}(2018)}]{rupke2018review}
{Rupke}, D. 2018, Galaxies, 6, 138, \dodoi{10.3390/galaxies6040138}

\bibitem[{{Rupke} {et~al.}(2005){Rupke}, {Veilleux}, \& {Sanders}}]{rupke2005outflowsI}
{Rupke}, D.~S., {Veilleux}, S., \& {Sanders}, D.~B. 2005, \apjs, 160, 87, \dodoi{10.1086/432886}

\bibitem[{{Rupke} {et~al.}(2019){Rupke}, {Coil}, {Geach}, {Tremonti}, {Diamond-Stanic}, {George}, {Hickox}, {Kepley}, {Leung}, {Moustakas}, {Rudnick}, \& {Sell}}]{rupke2019100}
{Rupke}, D. S.~N., {Coil}, A., {Geach}, J.~E., {et~al.} 2019, \nat, 574, 643, \dodoi{10.1038/s41586-019-1686-1}

\bibitem[{{Sato} {et~al.}(2009){Sato}, {Martin}, {Noeske}, {Koo}, \& {Lotz}}]{sato2009aegis}
{Sato}, T., {Martin}, C.~L., {Noeske}, K.~G., {Koo}, D.~C., \& {Lotz}, J.~M. 2009, \apj, 696, 214, \dodoi{10.1088/0004-637X/696/1/214}

\bibitem[{{Schaye} {et~al.}(2015){Schaye}, {Crain}, {Bower}, {Furlong}, {Schaller}, {Theuns}, {Dalla Vecchia}, {Frenk}, {McCarthy}, {Helly}, {Jenkins}, {Rosas-Guevara}, {White}, {Baes}, {Booth}, {Camps}, {Navarro}, {Qu}, {Rahmati}, {Sawala}, {Thomas}, \& {Trayford}}]{Schaye2015EAGLEsimulation}
{Schaye}, J., {Crain}, R.~A., {Bower}, R.~G., {et~al.} 2015, \mnras, 446, 521, \dodoi{10.1093/mnras/stu2058}

\bibitem[{{Shaban} {et~al.}(2022){Shaban}, {Bordoloi}, {Chisholm}, {Sharma}, {Sharon}, {Rigby}, {Gladders}, {Bayliss}, {Barrientos}, {Lopez}, {Tejos}, {Ledoux}, \& {Florian}}]{Shaban202230kpc}
{Shaban}, A., {Bordoloi}, R., {Chisholm}, J., {et~al.} 2022, \apj, 936, 77, \dodoi{10.3847/1538-4357/ac7c65}

\bibitem[{{Shaban} {et~al.}(2023){Shaban}, {Bordoloi}, {Chisholm}, {Rigby}, {Sharma}, {Sharon}, {Tejos}, {Bayliss}, {Barrientos}, {Lopez}, {Ledoux}, {Gladders}, \& {Florian}}]{shaban2023dissecting}
---. 2023, \mnras, 526, 6297, \dodoi{10.1093/mnras/stad3004}

\bibitem[{{Sharon} {et~al.}(2020){Sharon}, {Bayliss}, {Dahle}, {Dunham}, {Florian}, {Gladders}, {Johnson}, {Mahler}, {Paterno-Mahler}, {Rigby}, {Whitaker}, {Akhshik}, {Koester}, {Murray}, {Remolina Gonz{\'a}lez}, \& {Wuyts}}]{sharon2020strong}
{Sharon}, K., {Bayliss}, M.~B., {Dahle}, H., {et~al.} 2020, \apjs, 247, 12, \dodoi{10.3847/1538-4365/ab5f13}

\bibitem[{{Somerville} \& {Dav{\'e}}(2015)}]{somerville2015physical}
{Somerville}, R.~S., \& {Dav{\'e}}, R. 2015, \araa, 53, 51, \dodoi{10.1146/annurev-astro-082812-140951}

\bibitem[{{Somerville} {et~al.}(2008){Somerville}, {Hopkins}, {Cox}, {Robertson}, \& {Hernquist}}]{Somerville2008Semi-analyticModel}
{Somerville}, R.~S., {Hopkins}, P.~F., {Cox}, T.~J., {Robertson}, B.~E., \& {Hernquist}, L. 2008, \mnras, 391, 481, \dodoi{10.1111/j.1365-2966.2008.13805.x}

\bibitem[{{Soto} {et~al.}(2016){Soto}, {Lilly}, {Bacon}, {Richard}, \& {Conseil}}]{soto2016zap}
{Soto}, K.~T., {Lilly}, S.~J., {Bacon}, R., {Richard}, J., \& {Conseil}, S. 2016, \mnras, 458, 3210, \dodoi{10.1093/mnras/stw474}

\bibitem[{{Speagle}(2020)}]{Speagle2020Dynesty}
{Speagle}, J.~S. 2020, \mnras, 493, 3132, \dodoi{10.1093/mnras/staa278}

\bibitem[{{Spilker} {et~al.}(2019){Spilker}, {Bezanson}, {Weiner}, {Whitaker}, \& {Williams}}]{Spilker2019qQuench}
{Spilker}, J.~S., {Bezanson}, R., {Weiner}, B.~J., {Whitaker}, K.~E., \& {Williams}, C.~C. 2019, \apj, 883, 81, \dodoi{10.3847/1538-4357/ab3804}

\bibitem[{{Stewart} {et~al.}(2013){Stewart}, {Brooks}, {Bullock}, {Diemand}, {Wadsley}, \& {Moustakas}}]{Stewart2013AngularMomentum}
{Stewart}, K.~R., {Brooks}, A.~M., {Bullock}, James S. aFIXnd~{Maller}, A.~H., {et~al.} 2013, \apj, 769, 74, \dodoi{10.1088/0004-637X/769/1/74}

\bibitem[{{Strickland} \& {Heckman}(2009)}]{Strickland2009SupernovaFeedbackM82}
{Strickland}, D.~K., \& {Heckman}, T.~M. 2009, \apj, 697, 2030, \dodoi{10.1088/0004-637X/697/2/2030}

\bibitem[{{Tejos} {et~al.}(2021){Tejos}, {L{\'o}pez}, {Ledoux}, {Fern{\'a}ndez-Figueroa}, {Rivas}, {Sharon}, {Johnston}, {Florian}, {D'Ago}, {Katsianis}, {Barrientos}, {Berg}, {Corro-Guerra}, {Hamel}, {Moya-Sierralta}, {Poudel}, {Rigby}, \& {Solimano}}]{Tejos2021TelltaleCGM}
{Tejos}, N., {L{\'o}pez}, S., {Ledoux}, C., {et~al.} 2021, \mnras, 507, 663, \dodoi{10.1093/mnras/stab2147}

\bibitem[{Thompson \& Heckman(2024)}]{thompson2024theory}
Thompson, T.~A., \& Heckman, T.~M. 2024, Annual Review of Astronomy and Astrophysics, 62

\bibitem[{{Thompson} {et~al.}(2016){Thompson}, {Quataert}, {Zhang}, \& {Weinberg}}]{Thompson2016AnOriginMultiphase}
{Thompson}, T.~A., {Quataert}, E., {Zhang}, D., \& {Weinberg}, D.~H. 2016, \mnras, 455, 1830, \dodoi{10.1093/mnras/stv2428}

\bibitem[{{Tonry} {et~al.}(2012){Tonry}, {Stubbs}, {Lykke}, {Doherty}, {Shivvers}, {Burgett}, {Chambers}, {Hodapp}, {Kaiser}, {Kudritzki}, {Magnier}, {Morgan}, {Price}, \& {Wainscoat}}]{Tonry2012PanSTARRS1Photo}
{Tonry}, J.~L., {Stubbs}, C.~W., {Lykke}, K.~R., {et~al.} 2012, \apj, 750, 99, \dodoi{10.1088/0004-637X/750/2/99}

\bibitem[{{Tremonti} {et~al.}(2004){Tremonti}, {Heckman}, {Kauffmann}, {Brinchmann}, {Charlot}, {White}, {Seibert}, {Peng}, {Schlegel}, {Uomoto}, {Fukugita}, \& {Brinkmann}}]{tremonti2004origin}
{Tremonti}, C.~A., {Heckman}, T.~M., {Kauffmann}, G., {et~al.} 2004, \apj, 613, 898, \dodoi{10.1086/423264}

\bibitem[{{Tumlinson} {et~al.}(2017){Tumlinson}, {Peeples}, \& {Werk}}]{tumlinson2017circumgalactic}
{Tumlinson}, J., {Peeples}, M.~S., \& {Werk}, J.~K. 2017, \araa, 55, 389, \dodoi{10.1146/annurev-astro-091916-055240}

\bibitem[{{Veilleux} {et~al.}(2005){Veilleux}, {Cecil}, \& {Bland-Hawthorn}}]{veilleux2005galactic}
{Veilleux}, S., {Cecil}, G., \& {Bland-Hawthorn}, J. 2005, \araa, 43, 769, \dodoi{10.1146/annurev.astro.43.072103.150610}

\bibitem[{{Veilleux} {et~al.}(2020){Veilleux}, {Maiolino}, {Bolatto}, \& {Aalto}}]{Veilleux2020CoolOutflows}
{Veilleux}, S., {Maiolino}, R., {Bolatto}, A.~D., \& {Aalto}, S. 2020, \aapr, 28, 2, \dodoi{10.1007/s00159-019-0121-9}

\bibitem[{{Vogelsberger} {et~al.}(2020){Vogelsberger}, {Marinacci}, {Torrey}, \& {Puchwein}}]{Vogelsberger2020CosmoSimGalaxy}
{Vogelsberger}, M., {Marinacci}, F., {Torrey}, P., \& {Puchwein}, E. 2020, Nature Reviews Physics, 2, 42, \dodoi{10.1038/s42254-019-0127-2}

\bibitem[{{Vogelsberger} {et~al.}(2014){Vogelsberger}, {Genel}, {Springel}, {Torrey}, {Sijacki}, {Xu}, {Snyder}, {Nelson}, \& {Hernquist}}]{Vogelsberger2014Illustris}
{Vogelsberger}, M., {Genel}, S., {Springel}, V., {et~al.} 2014, \mnras, 444, 1518, \dodoi{10.1093/mnras/stu1536}

\bibitem[{{Weilbacher} {et~al.}(2016){Weilbacher}, {Streicher}, \& {Palsa}}]{Weilbacher2016MUSEPipeline}
{Weilbacher}, P.~M., {Streicher}, O., \& {Palsa}, R. 2016, {MUSE-DRP: MUSE Data Reduction Pipeline}, Astrophysics Source Code Library, record ascl:1610.004

\bibitem[{{Weilbacher} {et~al.}(2020){Weilbacher}, {Palsa}, {Streicher}, {Bacon}, {Urrutia}, {Wisotzki}, {Conseil}, {Husemann}, {Jarno}, {Kelz}, {P{\'e}contal-Rousset}, {Richard}, {Roth}, {Selman}, \& {Vernet}}]{Weilbacher2020MUSEPipeline}
{Weilbacher}, P.~M., {Palsa}, R., {Streicher}, O., {et~al.} 2020, \aap, 641, A28, \dodoi{10.1051/0004-6361/202037855}

\bibitem[{{Werk} {et~al.}(2014){Werk}, {Prochaska}, {Tumlinson}, {Peeples}, {Tripp}, {Fox}, {Lehner}, {Thom}, {O'Meara}, {Ford}, {Bordoloi}, {Katz}, {Tejos}, {Oppenheimer}, {Dav{\'e}}, \& {Weinberg}}]{Werk2014COSHalosCGM}
{Werk}, J.~K., {Prochaska}, J.~X., {Tumlinson}, J., {et~al.} 2014, \apj, 792, 8, \dodoi{10.1088/0004-637X/792/1/8}

\bibitem[{{Xu} {et~al.}(2022){Xu}, {Heckman}, {Henry}, {Berg}, {Chisholm}, {James}, {Martin}, {Stark}, {Aloisi}, {Amor{\'\i}n}, {Arellano-C{\'o}rdova}, {Bordoloi}, {Charlot}, {Chen}, {Hayes}, {Mingozzi}, {Sugahara}, {Kewley}, {Ouchi}, {Scarlata}, \& {Steidel}}]{Xu2022CLASSYIII}
{Xu}, X., {Heckman}, T., {Henry}, A., {et~al.} 2022, \apj, 933, 222, \dodoi{10.3847/1538-4357/ac6d56}

\bibitem[{{Xu} {et~al.}(2023){Xu}, {Ouchi}, {Nakajima}, {Harikane}, {Isobe}, {Ono}, {Umeda}, \& {Zhang}}]{Xu2023OutflowsJWST}
{Xu}, Y., {Ouchi}, M., {Nakajima}, K., {et~al.} 2023, arXiv e-prints, arXiv:2310.06614, \dodoi{10.48550/arXiv.2310.06614}

\bibitem[{{Zabl} {et~al.}(2021){Zabl}, {Bouch{\'e}}, {Wisotzki}, {Schaye}, {Leclercq}, {Garel}, {Wendt}, {Schroetter}, {Muzahid}, {Cantalupo}, {Contini}, {Bacon}, {Brinchmann}, \& {Richard}}]{zabl2021muse}
{Zabl}, J., {Bouch{\'e}}, N.~F., {Wisotzki}, L., {et~al.} 2021, \mnras, \dodoi{10.1093/mnras/stab2165}

\bibitem[{{Zaritsky} {et~al.}(1994){Zaritsky}, {Kennicutt}, \& {Huchra}}]{Zaritsky1994metallicity}
{Zaritsky}, D., {Kennicutt}, Robert~C., J., \& {Huchra}, J.~P. 1994, \apj, 420, 87, \dodoi{10.1086/173544}

\end{thebibliography}
\bibliographystyle{aasjournal}

\end{document}